\documentclass[letterpaper]{article} 

\usepackage[submission]{aaai24}  
\usepackage{times}  
\usepackage{helvet}  
\usepackage{courier}  
\usepackage[hyphens]{url}  
\usepackage{graphicx} 
\usepackage{natbib}  
\usepackage{caption} 
\usepackage[dvipsnames]{xcolor}
\usepackage{pgfplots}
\usepackage{subfigure}
\usepackage{algpseudocode}
\usepackage{url}
\usepackage{tabularray}
\usepackage{amsmath} 
\usepackage{amssymb}
\usepackage{booktabs}
\usepackage{multirow}
\usepackage{appendix}
\usepgfplotslibrary{colormaps}
\usepackage{algorithm}
\usepackage{algorithmicx}
\usepackage{algpseudocode}

\usepackage{newfloat}
\usepackage{listings}
\DeclareCaptionStyle{ruled}{labelfont=normalfont,labelsep=colon,strut=off} 
\floatstyle{ruled}
\newfloat{listing}{tb}{lst}{}
\floatname{listing}{Listing}
\title{Balancing Augmentation with Edge Utility Filter for Signed GNNs }
\author{
Ke-Jia CHEN\textsuperscript{\rm 1}, 
Yaming JI\textsuperscript{\rm 1}, 
Youran XU\textsuperscript{\rm 1}, 
Chuhan XU\textsuperscript{\rm 1}
}
\affiliations{
\textsuperscript{\rm 1}Nanjing University of Posts and Telecommunications, China\\

}

\usepackage{bibentry}

\begin{document}

\maketitle

\begin{abstract}
	Signed graph neural networks (SGNNs) has recently drawn more attention as many real-world networks are signed networks containing two types of edges: positive and negative. The existence of negative edges affects the SGNN robustness on two aspects. One is the semantic imbalance as the negative edges are usually hard to obtain though they can provide potentially useful information. The other is the structural unbalance, e.g. unbalanced triangles, an indication of incompatible relationship among nodes. In this paper, we propose a balancing augmentation method to address the above two aspects for SGNNs. Firstly, the utility of each negative edge is measured by calculating its occurrence in unbalanced structures. Secondly, the original signed graph is selectively augmented with the use of (1) an edge perturbation regulator to balance the number of positive and negative edges and to determine the ratio of perturbed edges to original edges and (2) an edge utility filter to remove the negative edges with low utility to make the graph structure more balanced. Finally, a SGNN is trained on the augmented graph which effectively explores the credible relationships. A detailed theoretical analysis is also conducted to prove the effectiveness of each module. Experiments on five real-world datasets in link prediction demonstrate that our method has the advantages of effectiveness and generalization and can significantly improve the performance of SGNN backbones.
\end{abstract}

\section{Introduction}

The relationships between nodes in many real networks (so- cial  networks, biological  networks,  etc.) can  express  not only positive semantics (e.g. trust, friends) but also neg- ative  semantics (e.g. distrust, hostility). This type of network is usually modeled as a signed network or signed graph containing both positive and negative edges. A series of signed graph learning methods \cite{7} 
\cite{109} \cite{110} have been developed in recent years, aiming to learn low-dimensional node representation through the leverage of graph structure and sociological theory. With the evolvement of deep learning technology, graph neural network (GNN) \cite{GNN} has become a powerful and popular graph learning tool to complete tasks such as node classification, node clustering and link prediction, which has recently been introduced in signed network embedding \cite{2}.

However, current GNN models are compromised when dealing with signed networks for the reasons below. Firstly, the message propagation mechanism cannot always work effectively as the edges of a signed network have two different semantics. Secondly, there is a serious imbalance in the semantic relationships between nodes as negative edges are more difficult to obtain though they can provide more potentially useful information. Thirdly, there are many unbalanced structures such as unbalanced triangles  \cite{112} indicating an incompatible relationship  among nodes which affects the robustness of signed GNNs \cite{104} \cite{106}. Overall, semantic imbalance and structural unbalance are two most important challenges in signed graph learning.

Graph data augmentation refers to slightly modifying existing graph data or generating its variants to increase the available training data. It can alleviate the structural sparsity caused by the power law distribution in the graph and the noise caused by data acquisition and transmission \cite{132}. In recent years, graph data augmentation has been widely used in unsigned graphs \cite{121}. In this paper, graph structural augmentation, a common type of graph data augmentation, is introduced in signed graph learning that can not only alleviate the network sparsity but also adjust the ratio of positive and negative edges to solve semantic imbalance. To our knowledge, this is the first work of using graph structural augmentation to implement a robust signed GNN. However, the unbalanced structures caused by the introduction of negative edges may degrade the performance of SGNN due to the propagation of noise \cite{13}. The  number  of  unbalanced structures may be further increased after graph structural augmentation since more negative edges are perturbed. Therefore, continuous attention should be paid to the impact of negative edges on structural balance when augmenting the graph.

Inspired  by the  structural  balance theory,  this paper  defines the  concept  of  edge \textit{utility} to quantify the contribution of a given negative edge to balanced structures.  Subsequently, the paper proposes a balancing augmentation method with edge utility filter for signed GNNs. The graph augmentation will be guided by the balanced structures as the negative edges contributing a large number of unbalanced structures will be filtered out as noise. An edge pertubation regulator is also used to adjust the proportion of perturbed edges and to balance their signs.

The main contributions of this paper are summarized as follows:

•  For the first time, we study the impact of negative edges to structural balance in SGNNs by proposing the concept of edge utility.

•  Graph   structural   augmentation   is   introduced   into SGNNs to alleviate both semantic imbalance and structural unbalance through an edge pertubation regulator and an edge utility filter, respectively.

•  Experiments show that the proposed  method can significantly improve the performance of SGNN and can be flexibly applied to various SGNN backbones.

\section{Related Work}
\subsection{Signed Network Embedding}
Signed graph embedding, commonly known as signed network embedding (SNE), is to embed nodes into a low-dimensional vector space while preserving the signed relationship between nodes in the original graph. SiNE \cite {2} applies deep learning to SNE for the first time, designing the embedding function based on social theory. At present, graph neural networks have become the mainstream method for SNE. SGCN \cite{7} first extends GCN \cite{8} to  signed  network  and  designs the positive  and  negative aggregators based on balance theory to aggregate the direct and indirect friend and foe information of each node. SSSNET \cite{100} introduces a new probabilistic balanced normalization loss reduction algorithm in GNN for embedding clustered signed network.
PbGCN \cite{10} combines balance theory and polarity information for message aggregation to capture edge  sign information on balanced structures. SigGAN \cite {12} is a GAN-based SNE framework, which generates edges while maintaining the structural balance of node triples. 

However, the above models are developed based on the assumption of structural balance, without considering the impact of the noise introduced by unbalanced structures.  Recently, Zhang et al. \cite{13} demonstrate that unbalanced triangles may bring noise and  thus propose RSGNN which modifies the loss to maintain the structural balance of the signed graph. Therefore, better utilization of structural balance in SNE is still worth studying.

Furthermore, positive and negative edges have completely different properties. The former exhibits high local clustering, small diameter and high transitivity, while the latter exhibits the opposite  \cite{101}.  In addition, the distribution of positive and negative edges in real signed networks is quite imbalanced, that is the latter is much sparser than the former. However, information contained from negative edges is usually more important for graph learning, e.g., they have a much higher impact on prediction accuracy than positive edges \cite{125}. Duan et al. \cite{14} also  indicate that an excellent negative sample can contribute diverse information that contrasts with a positive sample. Currently, little work has been developed to better learn negative edges despite their signficance, which is the gap this paper strives to bridge.

\subsection{Graph Structural Augmentation}
Graph structural augmentation refers to modifying the original graph structure or generating a new graph structure to augment graph data. Existing methods can be categorized into node-level \cite{15}, edge-level \cite{35}, sub-graph-level \cite{17}, graph-level \cite{22} and so on. Among them, edge-level augmentation is the most effective method, including edge removal, edge addition and the combination of both.

DropMessage \cite{103} provides a unified framework for most existing random dropping methods. Many graph contrast learning methods, such as SGCL \cite{35}, generate a contrastive graph view by randomly adding and removing edges, enhancing the robustness and generalization of the model. Although random edge augmentation shows certain effects, importance-based augmentation can better improve the model performance \cite{118}. The probability of an edge can be weighted according to heuristics like node centrality \cite{24} and node similarity \cite{25} to achieve selective edge augmentation.

With the development of GNN, some effective edge-level augmentation methods based on edge formation probability have recently emerged such as graph diffusion \cite{117} and graph autoencoder (GAE) \cite{32}. Ling et al. \cite{119} propose the Graphair method to learn fair representations. The augmentation includes node feature masking and edge perturbance through inner product decoding and sampling. AdaEdge \cite{102} adjusts the topology of the graph by removing inter-class edges and adding intra-class edges based on the predicted confidence of the edges. GraphAug \cite{120} calculates the probability of  edge perturbance using MLP  to generate augmented graphs that maintain label invariance.

However,  there  is  currently  a  lack  of  research  on graph augmentation for specific types of graphs, so this paper studies the signed graph augmentation strategy to explore useful negative relationships.

\begin{figure*}
	\centering
	\includegraphics[width=180mm]{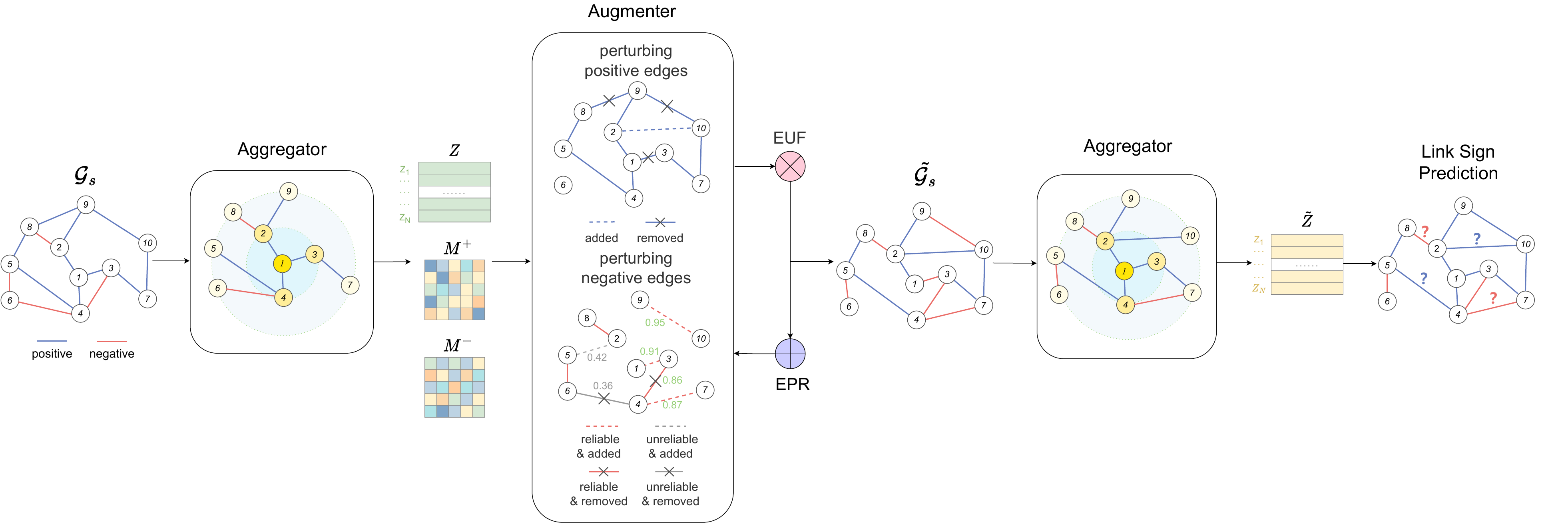}
	\centering\textbf{Figure 1:}~The framework of SiGAug.
	\label{fig1:framework}
\end{figure*}

\section{Preliminaries}

\textbf{Definition 1 Signed network.}
A signed network can be represented as a signed graph $\small  \mathcal{G}_{s} = (\mathcal{V} , \varepsilon ^{+} , \varepsilon ^{-} ) $, where $\small  \mathcal{V}=\left\{v_1,v_2,...,v_N\right\} $ is the set of $\small N$ nodes, $ \varepsilon ^{+}$ and $ \varepsilon ^{-}$ denote the  set  of positive and negative  edges in $\small  \mathcal{G}_{s}$, respectively. The adjacency matrix $\small  \mathcal{A}\:\in\:R^{N\times N} $ represents the connections in  $\small  \mathcal{G}_{s}$ , where  $\small  \mathit{\mathcal{A}_{ij} } $ is the entry $(i,j)$ of $\small \mathcal{A}$. If $\small  \mathit{\mathcal{A}_{ij} }=1 $, the edge $\mathit{e_{ij} }= \left<{{{v}}{}_{{i}}},{{v}}{}_{j{}}\right>  \in \varepsilon ^{+}$; if $ \small \mathit{\mathcal{A}_{ij} }=-1 $, the edge $\mathit{e_{ij} }= \left<{{{v}}{}_{{i}}},{{v}}{}_{j{}}\right>  \in \varepsilon ^{-}$; if $ \small \mathit{\mathcal{A}_{ij} }=0 $, there is no edge between $ {{v}}{}_{{i}} $ and $ {{v}}{}_{{j}} $.

In the study of signed networks, social balance theory \cite{40} is often adopted. On this basis, this paper defines two new concepts namely, $positive/negative\;path \;$and $balanced/unbalanced\;cycle$.

\textbf{Definition 2 Positive/negative path.}
Suppose $\small  \mathcal{P}_{ij}=\left\{{e_{i1},e_{12},...,e_{nj}}\right\} $($\small \left | \mathcal{P}_{ij}   \right |=n$) is a path in $\small  \mathcal{G}_{s} $ consisting of $n$ signed edges, where the starting and ending nodes are $ {{v}}{}_{{i}} $ and $ {{v}}{}_{{j}} $ respectively. $ \small SP(\cdot)$ is the path sign function, defined as the cumulative product of all edge signs on the path.
\begin{equation}
	\small
	\setlength{\abovedisplayskip}{0.5pt}
	\begin{aligned}
		SP(\mathcal{P}_{ij})=\prod_{{e}_k\in \mathcal{P}_{ij}}^n{{{e}_k}}
	\end{aligned}
	\setlength{\belowdisplayskip}{0.5pt}
\end{equation}
$\small SP(\mathcal{P}_{ij})=1$, if $\small \mathcal{P}_{ij}$ is a positive path with even negative edges; $\small SP(\mathcal{P}_{ij})=-1$, if $\small \mathcal{P}_{ij}$ is a negative path with odd negative edges.

\textbf{Definition 3 Balanced/unbalanced cycle.}
If the path $\small \mathcal{P}_{ij}=\left\{{e_{i1},e_{12},...,e_{nj}}\right\} $ has an edge $e_{ij}$ between $v_{i}$ and $v_{j}$, there will be a closed path $\small \mathcal{O}_{ij}=\left\{{e_{i1},e_{12},...,e_{nj}, e_{ji}}\right\} $, which is named as a \textit{cycle}. The cycle $\small \mathcal{O}_{ij}$ is a balanced (i.e. $\small SP(\mathcal{O}_{ij})=1$) if it contains even negative edges; otherwise, it is an unbalance cycle (i.e. $\small SP(\mathcal{O}_{ij})=-1$). For example, the balanced triangle is a special case of balanced cycle.

\textbf{Definition 4 Degree of edge utility.}
This is an original concept proposed in this paper. Any edge $e_{ij}$ in $\small  \mathcal{G}_{s} $ may be involved in two kinds of cycles: balanced and unbalanced.  Let  $\small \mathcal{Q}_{ij}=\left\{{\mathcal{O}_{ij}^{k}}\right\}(\left|\mathcal{O}_{ij}\right|\leq n)$ represent the set of all cycles with the length not greater than $n$ where $e_{ij}$  is located. $\small \mathcal{Q}_{ij}^{B}=\left\{{\mathcal{O}_{ij}^{k}}\middle|{SP\left(\mathcal{O}_{ij}^{k}\right)}=1 \right\} $ is the set of all balanced cycles where $e_{ij}$ is located. The degree of edge utility ${Utl}(e_{ij})$ is defined as the ratio of $\small \mathcal{Q}_{ij}^{B}$ to $\small \mathcal{Q}_{ij}$. The more balanced structures $e_{ij}$  contributes, the more valuable it becomes.

\textbf{Problem 1 Signed graph embedding.}
Given a signed graph $ \small \mathcal{G}_{s} $ with an adjacency matrix $\small  \mathcal{A}\:\in\:R^{N\times N} $, and a node feature matrix $\small \mathcal{X}\in R^{N\times D}$  (if any), signed graph embedding is to learn a mapping function:
\begin{equation}
	\small
	\setlength{\abovedisplayskip}{0.5pt}
	\begin{aligned}
		f_{\theta } :\mathcal{A},\mathcal{X}\to \mathcal{Z}\left ( \mathcal{Z}\in R^{N\times d}  \right ) 
	\end{aligned}
	\setlength{\belowdisplayskip}{0.5pt}
\end{equation}
where $\small  d\left ( d\ll D \right ) $ is the dimension of embedding vectors.

\textbf{Problem 2 Signed graph structural augmentation.}
Signed graph structural augmentation refers to perturbing nodes or signed edges in a signed graph $\small \mathcal{G}_{s}$. In this paper, only the perturbation of edges is considered.  Note that the signs of edges are also perturbed. The augmentation can be formalized as a function:
\begin{equation}
	\small
	\setlength{\abovedisplayskip}{0.5pt}
	\begin{aligned}
		f_{Aug} :\left ( \mathcal{G}_{s} ,\mathcal{A},\mathcal{X} \right ) \to \left ( \tilde{\mathcal{G}_{s} },\tilde{\mathcal{A}} ,\tilde{\mathcal{X}}   \right ) 
	\end{aligned}
	\setlength{\belowdisplayskip}{0.5pt}
\end{equation}
where $\small \tilde{\mathcal{G}_{s}}$  is the augmented graph, $\small \tilde{\mathcal{A}}$ is the adjacency matrix of $\small \tilde{\mathcal{G}_{s}}$ and $\small \tilde{X}$ is the node feature matrix of $\small \tilde{\mathcal{G}_{s}}$.

\section{Method}
This section introduces the proposed balancing augmentation method with edge utility filter for signed GNNs  (SiGAug).   Firstly,  an  edge  utility filter  is  constructed  which  defines  the utility of an edge by counting its frequency in balanced structures; secondly, a selective augmentation is performed on the original signed graph to explore the potential relationships (especially the negative edges with high utility); finally, the SGNN is trained on the augmented graph with balanced structures and semantics.

The overall framework of SiGAug is shown in Figure 1, which includes three modules: Edge Utility Filter (EUF), Aggregator, Augmenter with EUF and Edge Perturbation Regulator (EPR).  Each module is described as follows.

\subsection{Edge Utility Filter (EUF)}
Edge utility filter  (EUF)  can retain the edges that  are in  more balanced  structures,  which are considered more valuable for learning (see Definition 4).  The utility of negative edges is especially calculated since they decide directly whether the structure is balanced or not. The EUF module will be used in edge-level graph structural augmentation.

Unbalanced structures have been shown to be highly correlated with noise and usually refer to unbalanced triangles \cite{13}. However, the number of triangles in real signed graphs is often insuﬀicient, so this paper additionally investigates cycles with path lengths greater than 3 to jointly compute the number of unbalanced structures.

Each edge may participate in two types of cycles: balanced and unbalanced. For all edges, $\small \mathcal{C}^{B} \left ( n \right ) \in \mathbb{R}^{N \times N} $  represents the counting matrix of balanced cycles with length $n$,  $\small  \mathcal{C}^{U} \left ( n \right ) \in \mathbb{R}^{N \times N} $  represents the counting matrix of unbalanced cycles with length $n$ and $\small \mathcal{C}\left({n}\right) \in \mathbb{R}^{N \times N} $ is the counting matrix of all cycles with length $n$.

The EUF module consists of three main steps:  (1) calculating the number of balanced, unbalanced, and total cycles for each edge; (2) calculating the utility of each negative edge; (3) filtering out negative edges with low utility.

In the first step, $\small \mathcal{C}^{B} \left ( n \right )$, $\small \mathcal{C}^{U} \left ( n \right )$, and $\small \mathcal{C}\left({n}\right)$ ($ n\:\in3,4,\cdots $) are calculated. Specifically, the adjacency matrix $\small \mathcal{A}$ is first separated into $\small \mathcal{A}^{+}$ containing only positive edges $\varepsilon ^{+}$ and $\small \mathcal{A}^{-}$  containing only negative edges $\varepsilon ^{-}$. When $n=3$, for any negative edge $e_{ij}$, the number of balanced triangles is the number of unbalanced paths with length 2 between $ {{v}}_{{i}} $ and $ {{v}}_{{j}} $, which contain 1 negative edge and 1 positive edge. The number of unbalanced triangles is the number of balanced paths with length 2 between $ {{v}}_{{i}} $ and $ {{v}}_{{j}} $, which contain 2 negative edges or 2 positive edges.
\begin{equation}
	\small
	\setlength{\abovedisplayskip}{0.5pt}
	\begin{aligned}
		{\mathcal{C}}_{{}}^{{B}}\left(3\right)_{}={{{{\left({\mathcal{A}^{{+}}}\cdot{\mathcal{A}^{-{}}}\right)+\left({\mathcal{A}^{-{}}}\cdot{\mathcal{A}^{+{}}}\right)}}}{}_{{}}},\forall e_{ij}\in\varepsilon ^{-}
	\end{aligned}
	\setlength{\belowdisplayskip}{0.5pt}
\end{equation}
\begin{equation}
	\small
	\setlength{\abovedisplayskip}{0.5pt}
	\begin{aligned} {\mathcal{C}}_{{}}^{U{}}\left(3\right)_{}={\left({\mathcal{A}^{{+}}}\right)^2+\left({\mathcal{A}^{-{}}}\right)^2},\forall e_{ij}\in\varepsilon ^{+}
	\end{aligned}
	\setlength{\belowdisplayskip}{0.5pt}
\end{equation}
When $ n\geq4 $, $\small \mathcal{C}^{B} \left ( n \right )$ and $\small  \mathcal{C}^{U} \left ( n \right )$ are defined as Eq. 6 and Eq. 7, respectively. The multiplication by $\small \mathcal{A}^{-}$ indicates that negative edges will change the balance while the multiplication by $\small \mathcal{A}^{+}$ indicates that positive edges will not change the balance. 
\setlength{\abovedisplayskip}{0.5pt}
\begin{equation}
	\setlength{\abovedisplayskip}{0.5pt}
	\small
	\begin{aligned}
		{\mathcal{C}}_{{}}^{{B}}\left(n\right){{={\mathcal{C}}_{{}}^{{B}}\left(n-1\right)\cdot \mathcal{A}}}^{{+}}+{\mathcal{C}}_{{}}^{U{}}\left(n-1\right)\cdot \mathcal{A}^{-{}} 
	\end{aligned}
	\setlength{\belowdisplayskip}{0.5pt}
\end{equation}
\begin{equation}
	\small
	\begin{aligned}
		{\mathcal{C}}_{}^{{U}}\left(n\right){{={\mathcal{C}}_{}^{{B}}\left(n-1\right)\cdot \mathcal{A}}}^{{-}}+{\mathcal{C}}_{}^{U{}}\left(n-1\right)\cdot \mathcal{A}^{+} 
	\end{aligned}
\end{equation}
\setlength{\belowdisplayskip}{0.5pt}
Finally, the matrix $\small \mathcal{C}\left({n}\right)$ is the sum of $\small \mathcal{C}^{B} \left ( n \right )$ and $ \small \mathcal{C}^{U} \left ( n \right )$ .
\begin{equation}
	\small
	\begin{aligned} {{\mathcal{C}_{}\left({n}\right)=\mathcal{C}}}_{}^{B}\left({n}\right)+{{\mathcal{C}}}_{{}}^{U}\left({n}\right) \; ( 3\leq n\leq\eta )
	\end{aligned}
\end{equation}
where $\eta$ is the maximum length of the cycle.

In the second step, $\small {Utl}(e_{ij})$ (Eq. 9) is calculated. Considering that the longer the cycle, the weaker the impact of the edge on global structural balance, this paper set $\eta$ to 4. In other words, only the cycles of triangles and quadrilaterals are counted and summed.
\begin{equation}
	\small
	\begin{aligned}
		{Utl}(e_{ij})={{\frac{\sum_{n=3}^\eta{\mathcal{C}_{ij}^{B}\left({n}\right)}_{}}{\sum_{n=3}^\eta{\mathcal{C}_{ij}^{{}}\left({n}\right)}_{}}}}  	
	\end{aligned}
\end{equation}

In the third step, if $\small {Utl}(e_{ij})< \mu $ ($ \mu $ is the utility threshold), the edge $e_{ij}$ is discarded like noise as it contributes more unbalanced structures.

\subsection{Aggregator}
In SGNNs, messages are propagated from both positive and negative paths so the center node can aggregate both types of information. The positive embedding matrix $\small \mathcal{Z}^{+}$ and the negative embedding matrix $\small \mathcal{Z}^{-}$ can be obtained through multi-layer aggregation, and the final embedding matrix $ \small \mathcal{Z}$ is the concatenation of both. The aggregator can be any aggregation-based SGNN model. A benchmark aggregation method can be found in Appendix
C.1.

\subsection{Augmenter with EUF and EPR}

The augmentation contains four steps:   (1) calculating probability matrices from positive and negative embedding matrices, respectively;  (2) perturbing positive and negative edges with maximum and minimum probability, respectively, and the perturbed negative edges are fed to the EUF for filtering; (3) balancing the ratio of perturbed positive and negative edges and adjusting the ratio of perturbed edges to original edges with an edge perturbation regulator; (4) Fusing the augmented positive and negative adjacency matrices.

\setlength{\parskip}{0pt}

In the first step, the matrices $\small \mathcal{\mathcal{Z}}^{+}$ and $\small \mathcal{Z}^{-}$ obtained from the aggregation module are fed into an edge predictor to compute two edge probability matrices $\small \mathcal{M}^{+} \in \mathbb{R}^{N\times N}$ and $\small \mathcal{M}^{-} \in \mathbb{R}^{N\times N}$, respectively, where $\small \mathcal{M}_{ij}$ denotes the probability to form a signed edge between ${v}_{i}$ and $ {v}_{j}$. we use cosine similarity as the edge predictor, so that the probability of positive edges is positively correlated with the similarity and the probability of negative edges is negatively correlated with the similarity, i.e. $\small \mathcal{M}^{+}=\mathcal{Z}^{+} {\mathcal{Z}^+}^T $, $\small \mathcal{M}^{-}=1/\left ( \mathcal{Z}^{-} {\mathcal{Z}^-}^T \right )$.

In the second step, the maximum and the minimum value in $\small \mathcal{M}^{+}$ and $\small \mathcal{M}^{-}$ are all identified and the corresponding edges $e_{max}^+$, $e_{min}^+$, $e_{max}^-$, $e_{min}^-$ are perturbed, i.e. $\small \tilde{\mathcal{A}}^{+} _{max} =1$, $\small \tilde{\mathcal{A}}^{+} _{min} =0$, $\small \tilde{\mathcal{A}}^{-} _{max} =-1$, $\small \tilde{\mathcal{A}}^{-} _{min} =0$. Each perturbed negative edge $e_{ij}$  is sent to the EUF. If $\small Utl(e_{ij}) \ge \mu$, the edge is  retained and added into the set of perturbed edges $\hat{\varepsilon}$.

In the third step, the EPR is invoked after each perturbation to control augmentation through
two hyperparameters, the ratio of perturbed positive and negative edges $\small  \vartheta = \left | \hat{\varepsilon} ^{+} \right |/{ \left | \hat{\varepsilon} ^{-} \right |}$ and the ratio of perturbed edges to original edges $\small  \delta= {\left | \\ \hat{\varepsilon} \right |}/{\left | \varepsilon ^{} \right |}$. The augmentation process stops once both reach a predefined threshold, yielding the augmented adjacency matrix $\small \tilde{\mathcal{A}}$; otherwise, it returns to the second step.

In the last step, the two augmented adjacency matrices $\small \tilde{\mathcal{A}}^{+}$ and $\small \tilde{\mathcal{A}}^{-}$ are fused to $\small \tilde{\mathcal{A}}$. The fusion strategy is: if both $\small \tilde{\mathcal{A}}^{+}_{ij}$ and $\small \tilde{\mathcal{A}}^{-}_{ij}$  are 0, $\small \tilde{\mathcal{A}}_{ij}=0$; if one of the two is not zero, $\small \tilde{\mathcal{A}}_{ij}$ is set to that non-zero value; if both $\small \tilde{\mathcal{A}}^{+}_{ij}$ and $\small \tilde{\mathcal{A}}^{-}_{ij}$ are not zero, the value of $\small \tilde{\mathcal{A}}_{ij}$ is decided by the comparison of $\small \mathcal{M}^{+}_{ij}$ with $\small \mathcal{M}^{-}_{ij}$, that is, if $\small \mathcal{M}^{+}_{ij}>\mathcal{M}^{-}_{ij}$, $\small \tilde{\mathcal{A}}_{ij}=1$, otherwise, $\small \tilde{\mathcal{A}}_{ij}=-1$.

\subsection{Training}
\setlength{\parskip}{0pt}
The augmented graph $\small \tilde{\mathcal{G}_{s}}$ is fed into the SGNN again for training, and new embedding matrices $\small \tilde{\mathcal{Z}}^{+}$ and $\small \tilde{\mathcal{Z}}^{-}$ are concatenated to obtain the final matrix $\small \tilde{\mathcal{Z}}$. A weighted multinomial logistic regression (MLG) classifier is used to train the whole model by performing the link prediction task on the original graph $ \small \mathcal{G}_{s} $. The overall loss function is formalized as:

\begin{equation}
	\small
	\setlength{\abovedisplayskip}{0.5pt}
	\begin{aligned}
		&-\frac{1}{E}\sum_{\substack{\left({{v}}{}_{{i,}}{{v}}{}_{{j,}}s\right)\\\in E}}{{\omega}}{}_{{s}\:}\log\frac{exp([\mathcal{Z}_{i}, \mathcal{Z}_{j}]\theta _{s}^{})}{\sum_{\substack{q \\ \in({+,-,?})}}{exp([\mathcal{Z}_{i}, \mathcal{Z}_{j}]\theta _{q}^{})}} \\&+\lambda \Bigg[ {\frac{1}{\mid{E}{}_{{\left(+,?\right)}}\mid}\sum_{\substack{\left({{v}}{}_{{i}},{{v}}{}_{{j}}\right),\left({{v}}{}_{{i}},{{v}}{}_{{k}}\right)\\\in{E}{}_{{\left(+,?\right)}}}}{\max\left(0,\left({{\left \| {{{{\mathcal{Z}}}{}_{{i}}}-{{\mathcal{Z}}}{}_{{j}}} \right \| }}_{{2}}^{{2}}-\left \| {{{{\mathcal{Z}}}{}_{{i}}}-{{\mathcal{Z}}}{}_{{k}}} \right \| _{{2}}^{{2}}\right)\right)}} \\&+{\frac{1}{\mid{E}{}_{{\left(-,?\right)}}\mid}\sum_{\substack{\left({{v}}{}_{{i}},{{v}}{}_{{k}}\right),\left({{v}}{}_{{i}},{{v}}{}_{{j}}\right)\\\in{E}{}_{{\left(-,?\right)}}}}{\max\left(0,\left({{\left \|{{{{\mathcal{Z}}}{}_{{i}}}-{{\mathcal{Z}}}{}_{k{}}}\right \|}}_{{2}}^{{2}}-\left \|{{{{\mathcal{Z}}}{}_{{i}}}-{{\mathcal{Z}}}{}_{j{}}}\right \|_{{2}}^{{2}}\right)\right)}}\Bigg]\\&+reg\left(\theta^W,\theta^{MLG}\right)
	\end{aligned}
	\setlength{\belowdisplayskip}{0.5pt}
\end{equation}

where $\small \theta ^{W}$ is the weight matrix of the SGNN model, $\small \theta ^{MLG} $ is the parameter of MLG; $\small s, q\in\left\{{+,-,?} \right\}$ are the relationships between $v_{i}$ , $v_{j}$, which are positive, negative, or no edge; $\small W$ is the prediction weights of model; $\small E$ is the set of node pairs and $\small  \mid{E}\mid$  is the number of $\small E$ and $\small E_{\left ( +  ,? \right ) }$, $\small E_{\left ( -  ,? \right ) }$ are the set of positive and negative edges respectively; $\small \lambda$ controls the contribution of the MLG loss to the overall loss; $reg$ is for parameter regularization.

\subsection{Theoretical Analysis}

\subsubsection{Edge Augmenter to  Alleviate Over-smoothing}

Like GNN, SGNN also suffers from over-smoothing problem. We prove that edge augmenter can optimize the graph topology and diversify the passing messages via perturbation using Shannon entropy \cite{130}. Assume that $ \left|{\varepsilon}\right| $ types of d-dimensional messages are propagated in SGNN. The overall Shannon entropy is defined as:
\begin{equation}
	\small 
	\begin{aligned}
		H\left({\mathcal{G}_s }\right)=\sum_i^{ \left|{\varepsilon}\right| }{-p_i\log p_i} \;,
	\end{aligned}
\end{equation}
where $\small H\left(\mathcal{G}_s\right)$ is the Shannon entropy of $\mathcal{G}_s$. By perturbing the edges at a rate of $\delta$, the Shannon entropy can be rewritten as an expectation:
\begin{equation}
	\small
	\begin{aligned}  E(H(\tilde{\mathcal{G}_s}))=-\delta\log\left(\delta\right)+\left(1-\delta\right)\sum_i^{ \left|{\varepsilon}\right| }{-p_i\log\left(\left(1-\delta\right)p_i\right)}
	\end{aligned}
\end{equation}
We can get: $\small E(H(\tilde{\mathcal{G}_s}))\geq H\left(\mathcal{G}_s\right)$ at the appropriate perturbance rate $\delta$. It indicates that the over-smoothing can be alleviated. A detailed analysis can be found in Appendix A.1.

\subsubsection{Edge Utility Filter for Denoising}
SGNN cannot learn proper representations of nodes from unbalanced triangles which has been proved by constructing 2-hop ego trees \cite{13}. Similarly, we prove that  discarding negative  edges in unbalanced structures can help SGNN learn more proper representations of nodes. The required deﬁnitions and theorems can be found in Appendix A.2.

We draw 2-hop-ego-tree (Figure 2) for two types of unbalanced triangles. For the unbalanced triangle (a), $ t_i$ and $ t_j$ are isomorphic and will be mapped to the same embedding (Theorem 1), i.e.,  $\small h_{i}= h_{j}$. Therefore,  $\small dist(h_{i},h_{j})\leq dist(h_{i},h_{k}) $ , which means that the nodes connected with negative edges are closer. According to Definition 5, $\small  {{h}}{}_{{i}},{{h}}{}_{{j}},{{h}}{}_{{k}} $ are not proper representations. If we delete $e_{ij}$ (see Figure 2(c)), there is no negative edge between $v_{i}$ and $v_{j}$,  now $\small  {{h}}{}_{{i}},{{h}}{}_{{j}},{{h}}{}_{{k}} $  become proper representations. For the unbalanced triangle (b), $\small h_{i}=h_{j}=h_{k}$ , $\small dist(h_{i},h_{j})=0  $ and $\small dist(h_{i},h_{k})=0$, so $ \small {{h}}{}_{{i}},{{h}}{}_{{j}},{{h}}{}_{{k}} $  are not proper representations. After deleting the negative edge $e_{ij}$, and $\small  dist(h_{i},h_{k})\ne0 $ , so $\small  {{h}}{}_{{i}},{{h}}{}_{{j}},{{h}}{}_{{k}} $  become proper representations.

\begin{figure}
	\centering
	\includegraphics[width=90mm]{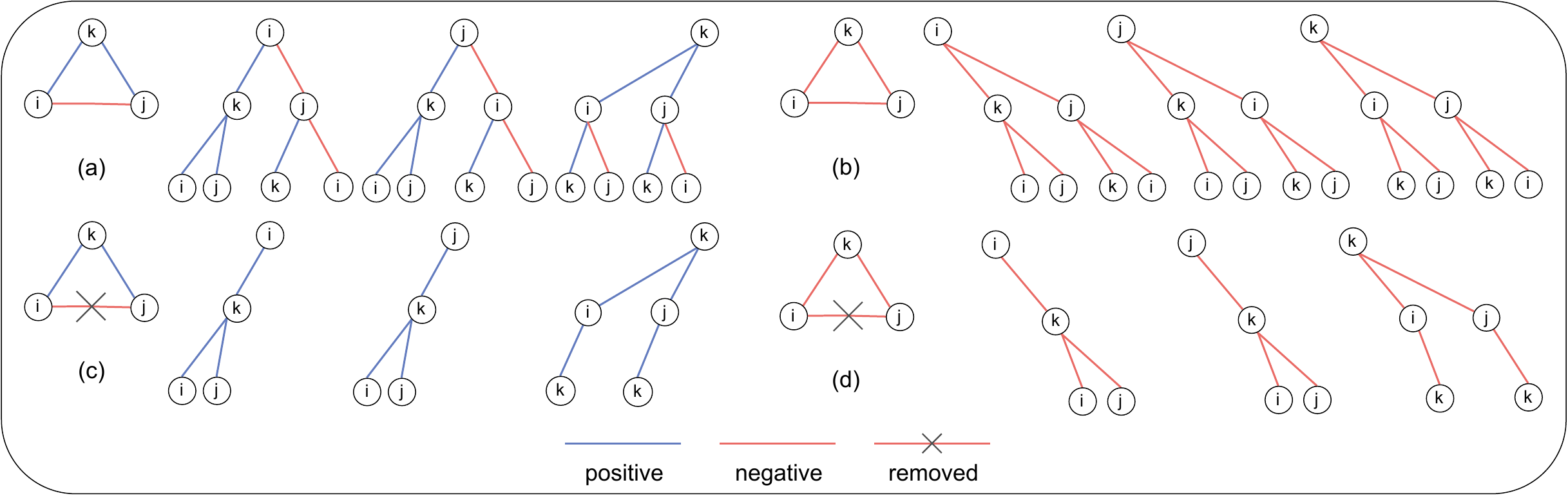}
	
	\centering\textbf{Figure 2:}~2-hop ego trees for unbalanced triangles.
	\label{fig2:filter}
\end{figure}

Similar conclusions can be obtained in unbalanced quadrilaterals (see Appendix A.2). Therefore, deleting negative edges in unbalanced cycles enables SGNN to learn better representations.

\subsubsection{Edge Regulator to Adjust Decision Boundary}

The distribution imbalance of  positive and negative  edges  can  bias the learning of SGNN. We prove that adjusting edge sampling frequency can improve SGNN’s performance. Taking the link prediction task as an example, from the loss function (Eq. 10), it is derived that the higher the sampling frequency, the larger the corresponding weight vector paradigm will be, which will affect the decision boundary (see Appendix A.3). We rewrite the decision boundary as a hyperplane:
\begin{equation}
	\small
	\begin{aligned}
		\mathcal{B}\left(i,j\right)=\left\{{\mathit{e}\in\:\varepsilon}\middle|{\left\|{w}_i\right\|_2\:\cos}\left({{\theta}}_{{e}}^{{i}}\right)={\left\|{w}_j\right\|_2\:\cos}\left({{\theta}}_{{e}}^{{j}}\right)\right\} 
	\end{aligned}
\end{equation} 
For a given graph, the direction of $w^+$ and $w^-$ is fixed after training a SGNN. As $\tau^+$ is usually larger than that of $\tau^-$, the paradigm $\small {\left \| w^{+} \right \|}_2 > {\left \| w^{-} \right \|}_2$ and the angle with the decision boundary $\small{{\theta}}^{{+}} >{{\theta}}^{{-}}$. It indicates that the decision boundary tends to be on negative edges, resulting in a smaller volume of feature space being allocated to negative edges. The edge regulator proposed in this paper tries to balance $\tau^+$ and  $\tau^-$ during graph augmentation, which can potentially adjust the decision boundary so as to alleviate the semantic unbalance problem in signed graphs. See Appendix A.3 for the detailed derivation process.

\begin{figure*}
	\begin{longtblr}[
		label = none,
		entry = none,
		]{
			width = \linewidth,
			colspec = {Q[87]Q[146]Q[137]Q[148]Q[137]Q[137]Q[144]},
			row{1} = {c},
			cell{1}{1} = {c=2}{0.232\linewidth},
			cell{2}{1} = {r=4}{c},
			cell{2}{3} = {c},
			cell{2}{4} = {c},
			cell{2}{5} = {c},
			cell{2}{6} = {c},
			cell{2}{7} = {c},
			cell{3}{3} = {c},
			cell{3}{4} = {c},
			cell{3}{5} = {c},
			cell{3}{6} = {c},
			cell{3}{7} = {c},
			cell{4}{3} = {c},
			cell{4}{4} = {c},
			cell{4}{5} = {c},
			cell{4}{6} = {c},
			cell{4}{7} = {c},
			cell{5}{3} = {c},
			cell{5}{4} = {c},
			cell{5}{5} = {c},
			cell{5}{6} = {c},
			cell{5}{7} = {c},
			cell{6}{1} = {r=4}{c},
			cell{6}{3} = {c},
			cell{6}{4} = {c},
			cell{6}{5} = {c},
			cell{6}{6} = {c},
			cell{6}{7} = {c},
			cell{7}{3} = {c},
			cell{7}{4} = {c},
			cell{7}{5} = {c},
			cell{7}{6} = {c},
			cell{7}{7} = {c},
			cell{8}{3} = {c},
			cell{8}{4} = {c},
			cell{8}{5} = {c},
			cell{8}{6} = {c},
			cell{8}{7} = {c},
			cell{9}{3} = {c},
			cell{9}{4} = {c},
			cell{9}{5} = {c},
			cell{9}{6} = {c},
			cell{9}{7} = {c},
			cell{10}{1} = {r=4}{c},
			cell{10}{3} = {c},
			cell{10}{4} = {c},
			cell{10}{5} = {c},
			cell{10}{6} = {c},
			cell{10}{7} = {c},
			cell{11}{3} = {c},
			cell{11}{4} = {c},
			cell{11}{5} = {c},
			cell{11}{6} = {c},
			cell{11}{7} = {c},
			cell{12}{3} = {c},
			cell{12}{4} = {c},
			cell{12}{5} = {c},
			cell{12}{6} = {c},
			cell{12}{7} = {c},
			cell{13}{3} = {c},
			cell{13}{4} = {c},
			cell{13}{5} = {c},
			cell{13}{6} = {c},
			cell{13}{7} = {c},
			hline{1-2,6,10,14} = {-}{},
		}
		Method &            & Bitcoin\_OTC & Bitcoin\_Alpha & Congress    & Chess       & Wiki\_Election \\
		\small SGCN & \small Original     & 0.812±0.007  & 0.714±0.005    & 0.521±0.060 & 0.668±0.006 & 0.763±0.011\\
		&\small DropMessage  & \underline{0.837±0.006}  & \underline{0.764±0.012}    & \underline{0.608±0.041} & \underline{0.681±0.004} & \underline{0.780±0.011}    \\
		&  \small GAug         & 0.602±0.001  & 0.624±0.002    & 0.561±0.082 & 0.562±0.001 & 0.636±0.016    \\
		&  \small   SiGAug(Ours)         &\pmb{0.947±0.003}  & \pmb{0.966±0.006}    & \pmb{0.898±0.074} & \pmb{0.889±0.012} & \pmb{0.920±0.005}    \\
		\small SNEA &\small Original     & 0.866±0.005  & 0.864±0.005    & 0.781±0.013 & 0.598±0.002 & \underline{0.817±0.003}    \\
		& \small DropMessage  & \underline{0.879±0.002}  & \underline{0.869±0.005}    & \underline{0.793±0.031} & 0.584±0.002 & 0.814±0.001    \\
		& \small GAug         & 0.668±0.008  & 0.671±0.007    & 0.750±0.019 & \underline{0.754±0.002} & 0.755±0.003    \\
		&\small SiGAug(Ours)         & \pmb{0.905±0.020}  &\pmb{0.933±0.015}    & \pmb{0.849±0.015} & \pmb{0.775±0.015} & \pmb{0.889±0.001}    \\
		\small RSGNN & \small Original    & \underline{0.792±0.012}  & \underline{0.781±0.018}    & 0.561±0.053 & \underline{0.609±0.004}  & 0.764±0.004    \\
		& \small DropMessage & 0.607±0.003  & 0.610±0.007    & 0.605±0.001 & 0.605±0.004 & 0.610±0.009    \\
		& \small GAug        & 0.728±0.021  & 0.763±0.015    & \underline{0.648±0.001} & 0.591±0.003 & \underline{0.789±0.002}    \\
		&\small SiGAug(Ours)        & \pmb{0.876±0.005}  & \pmb{0.908±0.006}    & \pmb{0.747±0.023} &    \pmb{0.823±0.003}   &  \pmb{0.813±0.003}           
	\end{longtblr}
	\centering\textbf{Table 1:}~~Link sign prediction performance on AUC.
\end{figure*}

\section{Experiment}

We experimentally evaluate the performance of the SiGAug method to answer the following questions:
\begin{itemize}
	\item  \textbf{Q1}: Can  SiGAug  mitigate  the  semantic  imbalance  and structural unbalance in signed graphs and thus improve the performance of SGNNs?
	\item \textbf{Q2}: Is SiGAug applicable to different SGNN backbones and can it improve their learning ability?
	\item \textbf{Q3}: How do different hyperparameters affect the effectiveness of SiGAug?
\end{itemize}

\subsection{Experimental Setup}

\noindent\textbf{Datasets}. We conducted experiments on five publicly available real signed graph datasets, Bitcoin\_OTC, Bitcoin\_Alpha, Congress, Chess, and Wiki\_Election, with varying degrees of semantic imbalance and structural unbalance.  The statistics of the datasets can be found in Appendix B.

\noindent\textbf{Backbone models.} We use three mainstream SGNNs as the backbone models: SGCN \cite{7}, RSGNN \cite{13} and SNEA \cite{110}. The augmentation methods DropMessage \cite{103}, GAug (the GAug-O version) \cite{121} and SiGAug are applied on the backbones, respectively. The first two performs the same augmentation on positive and negative edges in signed graphs.  We do not compare with unsigned GNNs since previous works on SGNNs \cite{104} has shown that they do not perform well in signed graphs.  More details on the backbone methods are given in Appendix C.1.

\noindent\textbf{Hyper-parameter settings.} Each backbone model adopts the reported optimal parameter settings. The embedding dimension of all methods is set to 64 and the epoch is set to 100 to achieve a fair comparison.  SiGAug is implemented by PyTorch, where the edge utility threshold $\mu$ (to control structural balance) in EUF is in the range of [0, 0.9], the ratio of positive and negative perturbed edges $\vartheta$ (to control semantic balance) is in the range of [0.2,1], and the ratio of perturbed edges to original edges $\delta$ (to control the perturbation rate) is in the range of [0,1].

\noindent\textbf{Task and evaluation metrics.} The training task is link sign prediction, predicting whether a future edge is positive or negative. For each dataset, 20\% of the existing edges are randomly selected as the test set, and the remaining 80\% as the training set. The evaluation metrics are the AUC and the binary average F1 score, which are commonly used for link prediction, and higher values indicate better performance. We report the average results of five runs.

\begin{figure*}[htb]
	\begin{longtblr}[
		label = none,
		entry = none,
		]{
			width = \linewidth,
			colspec = {Q[162]Q[148]Q[160]Q[148]Q[148]Q[156]},
			column{1} = {c},
			cell{1}{2} = {c},
			cell{1}{3} = {c},
			cell{1}{4} = {c},
			cell{1}{5} = {c},
			cell{1}{6} = {c},
			cell{2}{2} = {c},
			cell{2}{3} = {c},
			cell{2}{4} = {c},
			cell{2}{5} = {c},
			cell{2}{6} = {c},
			cell{3}{2} = {c},
			cell{3}{3} = {c},
			cell{3}{4} = {c},
			cell{3}{5} = {c},
			cell{3}{6} = {c},
			cell{4}{2} = {c},
			cell{4}{3} = {c},
			cell{4}{4} = {c},
			cell{4}{5} = {c},
			cell{4}{6} = {c},
			cell{5}{2} = {c},
			cell{5}{3} = {c},
			cell{5}{4} = {c},
			cell{5}{5} = {c},
			cell{5}{6} = {c},
			hline{1-2,6} = {-}{},
		}
		Method       & Bitcoin\_OTC & Bitcoin\_Alpha & Congress    & Chess       & Wiki\_Election \\
		\small Original     & 0.457±0.012  & 0.251±0.020    & 0.342±0.025 & 0.433±0.013 & 0.512±0.017    \\
		\small DropMessage  & 0.466±0.009  & 0.263±0.005    & 0.344±0.016 & 0.432±0.007 & 0.516±0.008    \\
		\small GAug         & 0.562±0.010  & 0.560±0.003    & 0.355±0.019 & 0.511±0.011 & 0.554±0.012    \\
		\small SiGAug(Ours) &  \pmb{0.886±0.007}  &  \pmb{0.914±0.011}    &  \pmb{0.744±0.028} &  \pmb{0.845±0.014} &  \pmb{0.821±0.023}  
	\end{longtblr}
	\centering\textbf{Table 2:}~~Negative link prediction performance on F1.
\end{figure*}

\subsection{Experimental Result and Analysis}

\begin{itemize}
	\item  \textbf{The   impact   of   SiGAug   on   the   performance   of SGNNs (Q1)}
\end{itemize}

\noindent \textit{Improvement on overall performance. }Table  1 demonstrates the AUC values of all augmentation methods applied on all backbones, and the F1-scores of all methods are listed in Appendix C.2 The results show that SiGAug achieves significant performance improvement on all datasets. Taking the AUC of SGCN as an example, SiGAug improves the performance of the original SGCN by 16\% to 72\%. It indicates that the two inherent problems of semantic imbalance and structural unbalance do hinder the SGNN model from learning the proper representations, which are greatly alleviated by our method. DropMessage and GAug are not consistently effective when applied to  signed  graphs.  For  example, GAug  degrades  the  performance of almost all backbones and DropMessage only shows a slight increase. It validates that the unsigned graph augmentation method is not directly applicable to signed graphs. Indiscriminate augmentation of positive and negative edges may exacerbate semantic imbalance and structural unbalance, causing a decrease in performance. In contrast, SiGAug obtains a significant performance improvement through an elaborate signed edge augmentation strategy, which explores as many credible negative edges as possible.

\noindent \textit{Improvements on negative edge prediction. }To further observe SiGAug’s ability of learning negative edges, we specifically summarize the negative edge prediction performance of all methods on five datasets (see Table 2). Taking SGCN as an example, the overall F1 improvement of SiGAug on the five datasets reaches from 11\% to 44\% (Appendix C.2), while the improvement on negative edge prediction is much more higher, reaching from 29\% to 264\%. The result is consistent with the theoretical analysis in 4.5.3 that the prediction performance of SGNN on small class (negative edges) can be greatly improved by balancing the decision boundary of positive and negative edges.

\begin{figure}[H]
	\centering
	\includegraphics[width=90mm]{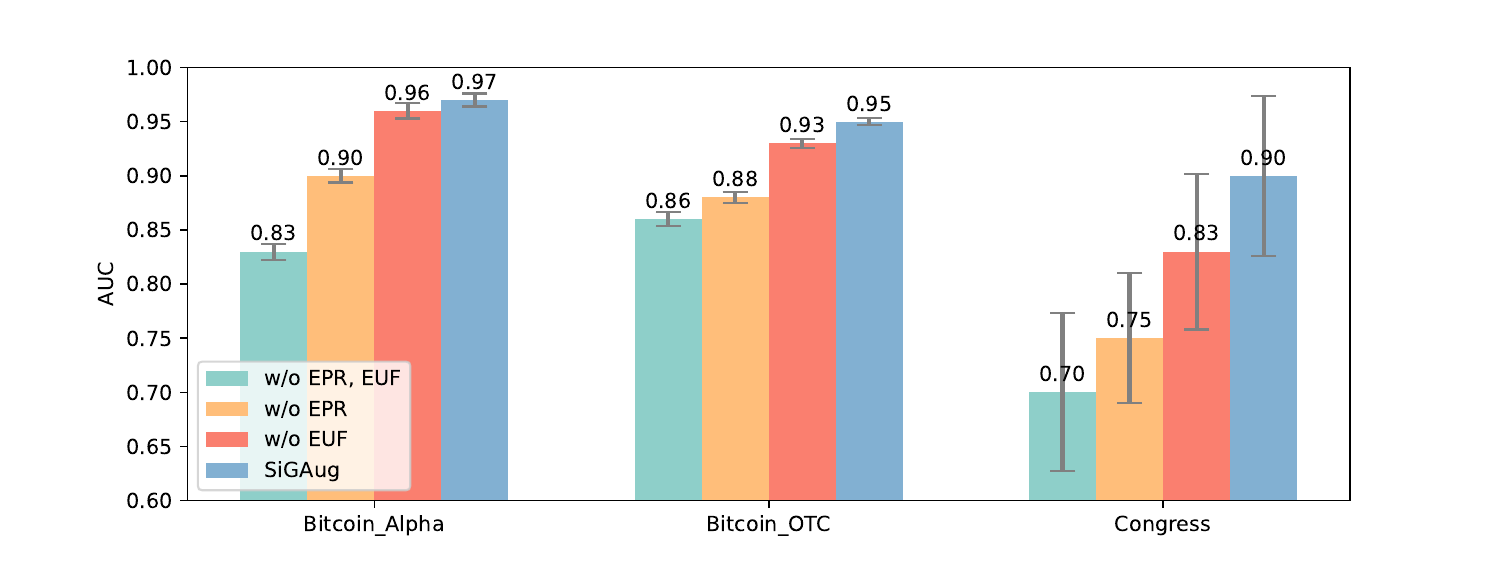}
	\centering{\textbf{Figure 3:}}~~The ablation study results of using different
	module.
	\label{fig1:ablation}
\end{figure}

\noindent \textit{Ablation Study. }We conduct ablation experiments to further study the effectiveness of EUF and EPR in SiGAug. Taking SGCN as the backbone, the results of each SiGAug variants on Bitcoin\_Alpha, Bitcoin\_OTC, Congress are shown in Figure 3. The results show that the improvement of AUC reaches 8.4\%, 2.3\%, 7.1\% respectively after adopting EUF, fully verifying that better representations are obtained by  denoising negative edges. The AUC of SiGAug increases by 7.7\%, 8.2\%, 20.0\% respectively compared to SiGAug-w/o EPR, which indicates that it is necessary to control the number and the signs of perturbed edges during the augmentation process. SiGAug-w/o EUF performs better than SiGAug-w/o EPR, suggesting that EPR has a greater impact on the augmenter. It is probably because most datasets have a high degree of global structural balance so EUF can not significantly alter the structure while EPR can obtain more accurate perturbed edges.

\begin{itemize}
	\item  \textbf{Effectiveness of SiGAug on different backbones (Q2)}
\end{itemize}

To answer Q2, we also report the AUC (see Table 1) and F1 performance (see Appendix C.3) of SiGAug on two other backbones: SNEA, RSGNN. The results show that SiGAug can be flexibly adapted to different SGNN backbones with different but stable improvements. DropMessage achieves suboptimal result in SGCN and SNEA, but not as good as GAug in RSGNN. The possible reason is that  RSGNN has potentially dropped noisy messages by modifying the loss, so the adding-edge methods like GAug and SiGAug is more effective. The use of SiGAug in RSGNN has the considerable improvement by 6\% to 35\% on all the datasets indicating that SiGAug can be effective on various signed graphs with different structural and semantic balance.

\begin{itemize}
	\item  \textbf{Influence of hyperparameters on SiGAug (Q3)}
\end{itemize}

\begin{figure} [!t]
	\centering 
	\setcounter{subfigure}{0}
	\subfigure[$\mu$-$\delta$, $\vartheta=1/9$]{
		\begin{minipage}[b]{0.46\linewidth}
			\pgfplotsset{width = 10cm,compat=1.5}
			\begin{tikzpicture}[thick,fill opacity=0.9,scale = 0.425]
				\begin{axis}[
					grid=major,major grid style={dashed}, 
					view={-45}{30},
					xmin=0, 
					xmax=1,    
					ymin=0, 
					ymax=1,
					zmin=0.9,
					enlarge z limits=0.2,  
					xlabel=$\delta$,
					ylabel=$\mu$ ,
					zlabel={\footnotesize AUC},
					ytick = {0.1,0.3,0.5,0.7,0.9},
					ztick={0.9,0.93,0.96},
					xlabel style={sloped like x axis},
					ylabel style={sloped like y axis},
					colormap/bone,
					]
					\addplot3 [surf,]
					coordinates {
						(0.2,0.9,0.909)  (0.2,0.7,0.919) (0.2,0.5,0.918) (0.2,0.3,0.907) (0.2,0.1,0.903) 
						
						(0.4,0.9,0.936) (0.4,0.7,0.951) (0.4,0.5,0.940) (0.4,0.3,0.944) (0.4,0.1,0.941) 
						
						(0.6,0.9,0.949)   (0.6,0.7,0.957) (0.6,0.5,0.953) (0.6,0.3,0.951) (0.6,0.1,0.951) 
						
						(0.8,0.9,0.955) (0.8,0.7,0.959) (0.8,0.5,0.965) (0.8,0.3,0.954) (0.8,0.1,0.953) 
						
						(1.0,0.9,0.961)   (1.0,0.7,0.955) (1.0,0.5,0.958) (1.0,0.3,0.962) (1.0,0.1,0.959) 
					};
				\end{axis}
			\end{tikzpicture}
		\end{minipage}
	}
	\subfigure[$\vartheta$-$\delta$, $\mu=0.7$]{
		\begin{minipage}[b]{0.46\linewidth}
			\pgfplotsset{width=10cm,compat=1.5}
			\begin{tikzpicture}[thick,fill opacity=0.9,scale = 0.425]
				\begin{axis}[
					grid=major,major grid style={dashed}, 
					view={-45}{30},
					xmin=0, 
					xmax=1,    
					xlabel=$\delta$,
					ylabel=$\vartheta$ ,
					zlabel={\footnotesize AUC},
					enlarge y limits=0.005,  
					xlabel style={sloped like x axis},
					ylabel style={sloped like y axis},
					symbolic y coords={8:2,6:4,4:6,2:8,1:9},
					ytick={8:2,6:4,4:6,2:8,1:9},
					colormap/gray
					]
					\addplot3 [surf,]
					coordinates {
						(0.2,1:9,0.927)  (0.2,2:8,0.890) (0.2,4:6,0.864) (0.2,6:4,0.851) (0.2,8:2,0.853) 
						
						(0.4,1:9,0.942) (0.4,2:8,0.936) (0.4,4:6,0.924) (0.4,6:4,0.907) (0.4,8:2,0.839) 
						
						(0.6,1:9,0.950)   (0.6,2:8,0.956) (0.6,4:6,0.922) (0.6,6:4,0.907) (0.6,8:2,0.910) 
						
						(0.8,1:9,0.967) (0.8,2:8,0.958) (0.8,4:6,0.953) (0.8,6:4,0.932) (0.8,8:2,0.896) 
						
						(1.0,1:9,0.966)   (1.0,2:8,0.959) (1.0,4:6,0.954) (1.0,6:4,0.934) (1.0,8:2,0.938) 
					};
				\end{axis}
			\end{tikzpicture}
		\end{minipage}
	}
	\\
	\centering\textbf{Figure 4:}~~The AUC of SiGAug-SGCN on Bitcoin\_Alpha when $\mu$, $\vartheta$ and $\delta$ vary.
	
\end{figure}

There are three core hyperparameters in SiGAug: the utility threshold $\mu$ in EUF, the ratio of perturbed positive and negative edges $\vartheta$ , and the ratio of perturbed edges to original edges $\delta$. $\mu$ varies from 0.1 to 0.9 in steps of 0.2; $\vartheta$ varies among 1:9, 2:8, 4:6, 6:4, 8:2 ; $\delta$ varies from 0.2 to 1 in steps of 0.2. Taking the AUC of SiGAug (SGCN as backbone) on Bitcoin\_Alpha as an example. As shown in Fig. 4(a), when $\mu$ is in [0.3, 0.7] and $\delta$ in [0.8, 1] , SiGAug performs the best. It indicates that perturbed more edges are helpful and meantime the threshold of the ﬁlter turns sensitive and need to be adjusted to retain more valuable negative edges. In Figure 4(b), AUC gets the optimal value when $\vartheta$ is 1/9, $\delta$ is in [0.8, 1]. In the Bitcoin\_Alpha dataset, the ratio of the original positive and negative edges is 10/1 (see Appendix B). The result is consistent with our expectation that  the model will perform better when the positive and negative edges reach a more balanced distribution after perturbation.

\begin{figure} [H]
	\centering 
	\setcounter{subfigure}{0}
	\subfigure[SNEA]{
		\begin{minipage}[b]{0.28\linewidth}
			\begin{tikzpicture}[scale = 0.31] 
				\begin{axis}[
					xmin=0, 
					xmax=1,    
					ymin=0.4, 
					ymax=1,
					enlarge x limits=0.1,  
					xlabel=$\delta$, 
					ylabel=AUC, 
					tick align=outside, 
					legend style={at={(1.7,1.3)},anchor=north,legend columns=-1} 
					]
					
					\addplot[,mark=*,violet] plot coordinates { 
						(0,0.864)
						(0.2,0.895)
						(0.4,0.911)
						(0.6,0.933)
						(0.8,0.916)
						(1.0,0.929)
					};
					
					\addlegendentry{\huge Bitcoin\_Alpha}
					
					
					\addplot[,mark=+,black] plot coordinates {
						(0,0.817)
						(0.2,0.835)
						(0.4,0.845)
						(0.6,0.871)
						(0.8,0.859)
						(1.0,0.885)
					};
					\addlegendentry{\huge Wiki\_election}
					
					\addplot[,mark=x,orange] plot coordinates {
						(0,0.598)
						(0.2,0.636)
						(0.4,0.719)
						(0.6,0.742)
						(0.8,0.724)
						(1.0,0.775)
					};
					\addlegendentry{\huge Chess}
					
					\addplot[,mark=triangle,cyan] plot coordinates {
						(0,0.781)
						(0.2,0.830)
						(0.4,0.845)
						(0.6,0.849)
						(0.8,0.837)
						(1.0,0.821)
					};
					\addlegendentry{ \huge Congress}
					
					\addplot[,mark=o,lime] plot coordinates {
						(0,0.866)
						(0.2,0.873)
						(0.4,0.875)
						(0.6,0.905)
						(0.8,0.879)
						(1.0,0.886)
					};
					\addlegendentry{\huge Bitcoin\_OTC}
				\end{axis}
			\end{tikzpicture}
		\end{minipage}
		\label{SNEA}
	}
	\subfigure[SGCN]{
		\begin{minipage}[b]{0.28\linewidth}
			\begin{tikzpicture}[scale =0.31] 
				\begin{axis}[
					xmin=0, 
					xmax=1,    
					ymin=0.4, 
					ymax=1,
					enlarge x limits=0.1,  
					xlabel=$\delta$, 
					ylabel=AUC, 
					tick align=outside, 
					legend style={at={(0.5,-0.2)},anchor=north} 
					]
					
					\addplot[,mark=*,violet] plot coordinates { 
						(0,0.714)
						(0.2,0.88)
						(0.4,0.937)
						(0.6,0.936)
						(0.8,0.948)
						(1.0,0.957)
					};
					
					
					
					\addplot[,mark=+,black] plot coordinates {
						(0,0.763)
						(0.2,0.844)
						(0.4,0.883)
						(0.6,0.911)
						(0.8,0.913)
						(1.0,0.894)
					};
					
					\addplot[,mark=x,orange] plot coordinates {
						(0,0.668)
						(0.2,0.779)
						(0.4,0.843)
						(0.6,0.866)
						(0.8,0.871)
						(1.0,0.88)
					};
					
					\addplot[,mark=triangle,cyan] plot coordinates {
						(0,0.521)
						(0.2,0.687)
						(0.4,0.777)
						(0.6,0.718)
						(0.8,0.766)
						(1.0,0.794)
					};
					
					\addplot[,mark=o,lime] plot coordinates {
						(0,0.812)
						(0.2,0.916)
						(0.4,0.920)
						(0.6,0.923)
						(0.8,0.947)
						(1.0,0.946)
					};
				\end{axis}
			\end{tikzpicture}
		\end{minipage}
		\label{fig:SGCN}
	}
	\subfigure[RSGNN]{
		\begin{minipage}[b]{0.28\linewidth}
			\begin{tikzpicture}[scale = 0.31] 
				\begin{axis}[
					xmin=0, 
					xmax=1,    
					ymin=0.4, 
					ymax=1,
					enlarge x limits=0.1,  
					xlabel=$\delta$, 
					ylabel=AUC, 
					tick align=outside, 
					legend style={at={(0.5,-0.2)},anchor=north} 
					]
					
					\addplot[,mark=*,violet] plot coordinates { 
						(0,0.781)
						(0.2,0.870)
						(0.4,0.892)
						(0.6,0.904)
						(0.8,0.895)
						(1.0,0.908)
					};
					
					
					
					\addplot[,mark=+,black] plot coordinates {
						(0,0.763)
						(0.2,0.812)
						(0.4,0.802)
						(0.6,0.787)
						(0.8,0.765)
						(1.0,0.742)
					};
					
					\addplot[,mark=x,orange] plot coordinates {
						(0,0.605)
						(0.2,0.638)
						(0.4,0.697)
						(0.6,0.733)
						(0.8,0.803)
						(1.0,0.824)
					};
					
					\addplot[,mark=triangle,cyan] plot coordinates {
						(0,0.561)
						(0.2,0.625)
						(0.4,0.725)
						(0.6,0.747)
						(0.8,0.658)
						(1.0,0.703)
					};
					
					\addplot[,mark=o,lime] plot coordinates {
						(0,0.792)
						(0.2,0.866)
						(0.4,0.862)
						(0.6,0.872)
						(0.8,0.876)
						(1.0,0.866)
					};
				\end{axis}
			\end{tikzpicture}
		\end{minipage}
		
		\label{RSGNN}
	}
	
	\centering\textbf{Figure 5:}~~The AUC of SiGAug when $\delta$ varies. 
\end{figure}

Figure 5 shows the fluctuation  of AUC for SiGAug with different backbones on all datasets as $\delta$  changes. The other two parameters are set to the optimal values, i.e., $\mu$=0.7 and $\vartheta$=1/9. When the pertubation rate $\delta$ is 0.6, SiGAug performs the best in most cases. It indicates that fully exploring the potential relationships in sparse signed graphs is very helpful for learning, but a higher pertubation rate may also introduce noise and lead to unstable performance. Therefore, the default value of $\mu$, $\vartheta$, $\delta$ in SiGAug is set to 0.7, 1/9, 0.6, respectively.

\subsection{Conclusion}
The existence of negative edges in signed graphs raises the problem of both semantic imbalance and structural unbalance, leading to unsatisfactory learning performance of SGNN. In this paper, we propose a signed graph augmentation method with edge utility filters and balancing perception, which fully explores potential negative edges and finally addresses the above challenges. We conduct detailed theoretical analysis and experiments to
demonstrate that our method  has a significant effect on improving the robustness of SGNN.
\bibliographystyle{aaai24}
\bibliography{aaai24}

\clearpage
\begin{appendices}
	
	\section*{Appendix}
	
	\section{A Theoretical Analysis}

	\section{A.1 Edge Augmenter to  Alleviate Over-smoothing}
	Shannon entropy is often used to measure the degree of confusion and diversity of variables and can be expressed as:
	$ H\left(X\right) E\left[{-\log p\left(X\right)}\right]=-\sum_{x\in\mathcal{X}}{p\left(x\right)}\log p\left(x\right) $
	
	where $\mathcal{X}$ represents the feature vector of the node and $\mathcal{F}$ represents all possible values of $\mathcal{X}$. Since there are $\left|{\varepsilon}\right| $ edges in $\mathcal{G}$, we assume that $ \left|{\varepsilon}\right| $ types of d-dimensional messages are propagated in SGNN. For the $i^{th}$ propagated messages, i.e., the messages delivered by the $i^{th}$ edge, which have been delivered  $T_i$ times by $n_i$ nodes, the Shannon entropy of the initial message is as follows:
	
	$$ H\left({\mathcal{G} }\right)=\sum_i^{ \left|{\varepsilon}\right| }{-p_i\log p_i} $$
	
	where $H\left(\mathcal{G}_s\right)$ is the Shannon entropy of $\mathcal{G}_s$, Now we perturb the edges at a perturbance rate of $\delta$ and the Shannon entropy can be expressed in the form of expectation:
	
	$$ E(H\left(SEA\right))=-\delta\log\left(\delta\right)+\left(1-\delta\right)\sum_i^{ \left|{\varepsilon}\right| }{-p_i\log\left(\left(1-\delta\right)p_i\right)} $$
	
	Since we have $ T_i\geq n_i $, at the appropriate perturbance rate $ \delta $ we get:
	$$ E(H\left(SEA\right))\geq H\left(\mathcal{G}\right)$$
	
	It proves that the over-smoothing problem in SGNN can be alleviated by perturbing the edges.
	
	\section{A.2 Edge Utility Filter for Denoising}
	
	\textbf{Definition 5. Proper representations of nodes. }Given a signed graph $ \mathcal{G} = (\mathcal{V} , \varepsilon ^{+} , \varepsilon ^{-} ) $, an SGNN model $ {{f}}{}_{{\theta}}:\mathcal{A},\mathcal{X}, \theta \to \mathcal{Z} $ and any non-negative distance metric $ dist:H\times H\rightarrow R^+ $, we call $ {{h}}{}_{{i}}={{f}}{}_{{\theta}}\left({{{v}}{}_{{i}}}\right) $ a proper representation of any node $ {{v}}{}_{{i}}\in V $ if the following conditions all satisfy:
	(a) There exist $ \zeta>0 $ such that for any $ {{v}}{}_{{j}}\in\varepsilon ^{-} $ and $ {{h}}{}_{{j}}={{f}}{}_{{\theta}}\left({{{v}}{}_{{j}}}\right) $, $ dist\left({{{{{{{{h}}{}_{{i}},h}}{}_{{j}}}}}}\right)>\zeta $;
	(b) For any $ {{v}}{}_{{j}}\in{{}}_{}^{{}}{{N}}_{{i}}^{{+}} $, $ {{v}}{}_{{k}}\in{{N}}_{{i}}^{{-}} $ and $ {{h}}{}_{{j}}={{f}}{}_{{\theta}}\left({{{v}}{}_{{j}}}\right) $,$ {{h}}{}_{{k}}={{f}}{}_{{\theta}}\left({{{v}}{}_{{k}}}\right) $ , $ dist\left({{{{{h}}{}_{{i}},h}}{}_{{j}}}\right)<dist\left({{{{{h}}{}_{{i}},h}}{}_k}\right) $.
	
	\textbf{Definition 6 k-hop ego-tree. } It is a k-level tree built from a root node $ {{v}}{}_{{i}} $ (level-0) in $  \mathcal{G}_{s} $. From any node $ {{v}}{}_{{j}} $ at level $ l\leq0 $, create a copy of each neighbor $ {{V}}{}_{{p}}\in{{N}}{}_{{j}} $ at level $l+1$ and connect  $ {{v}}{}_{{j}} $ and  $ {{v}}{}_{{p}} $ with a new tree edge with sign.
	
	\textbf{Theorem 1. } Suppose two ego-trees $ t_1 $ and $ t_2 $ are isomorphic. An SGNN applied to $ t_1$ and $ t_2$ will produce the same node embedding for the roots of $ t_1 $ and $ t_2 $.
	
	\begin{figure*}
		\centering
		\includegraphics[width=1\linewidth]{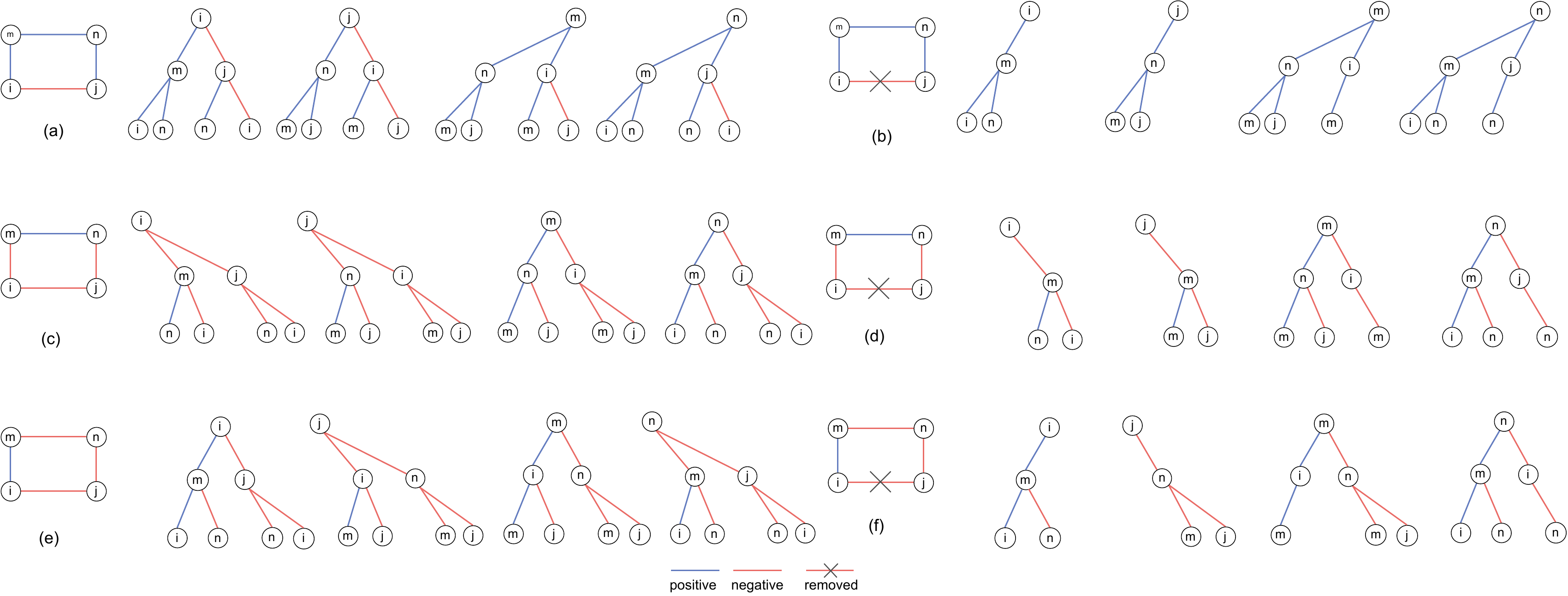}
		\centering\textbf{Figure 6:}~2-hop ego trees for unbalanced quadrilateral.
	\end{figure*}
	
	Similarly, for an unbalanced quadrilateral (a), $ t_{i} $ and $t_j$ are isomorphic and will be mapped to the same embedding (Theorem 1). i.e. , $h_{i}=h_{j}$, so that $ dist\left(h_{i},h_{j}\right) \leq dist\left(h_{i},h_{m}\right) $ ,$dist\left(h_{i},h_{j}\right) \leq dist\left(h_{i},h_{m}\right)$,  which means that the representation of the negative edge is closer than that of the positive edge, and $h_{i},h_{j},h_{m},h_{n}$ is not an appropriate representation according to Definition 5. If $e_{ij}$ is deleted, there is no negative edge between $v_{i}$ and $v_{j}$, $h_{i},h_{j},h_{m},h_{n}$ become proper representations.
	
	For unbalanced quadrilateral (c), $h_{i}=h_{j}$, $dist\left(h_{i},h_{j}\right)=0$, if $e_{ij}$ is deleted, there are no negative edges between $v_{i}$ and $v_{j}$, and $h_{i},h_{j},h_{m},h_{n}$ become proper representations. Similarly, for unbalanced quadrilateral (e), $h_{j}=h_{n}$, $dist\left(h_{j},h_{n}\right)=0$, if $e_{ij}$ is deleted, there are no negative edges between $v_{i}$ and $v_{j}$, $h_{i},h_{j},h_{m},h_{n}$ becomes the proper representation.

	\section{A.3 Edge Regulator to Adjust Decision Boundary}
	
	Taking the link prediction task as an example, the training data is the set of edges $ \varepsilon={{\left\{{\left({e},{{y}}{}\right)}\right\}}}  (\varepsilon^{+} \cup \varepsilon^{-}) $, where the labeling space $ y\in \left\{{1,-1}\right\} $, i.e., positive edges are labeled as 1 and negative edges are labeled as -1. $ f(\cdot) $ is the edge feature mapping function, and for any edge $\mathit{e }\in\:\varepsilon$, $ f(e) $ is its feature vector, which can be the arithmetic calculation of node embeddings. Assuming the multinomial logistic regression (MLG) is used as the classifier:
	$$ l(e)=W^Tf\left({e}\right)=\left[{{{w^+}}^Tf\left({e}\right);{{w^-}}^Tf\left({e}\right)}\right] $$
	where $w^{+} $,  $w^{-}$  are weight vector for positive and negative edges. The output logit vector $ l(e)\in\ \mathbb{R}^{2}$ is converted to a probability vector $ r(e)$ through the objective function softmax  $ r(e)=softmax(l(e))$ and the loss function is defined as:
	$$ \mathcal{L}\left(\varepsilon\right)=\frac{1}{\left|{\varepsilon}\right|}\sum_{e\in\varepsilon}{\ell\left(e,{{y}}{}_{{}}\right)}=\frac{\tau^+}{\left|{\varepsilon}\right|}\mathcal{L}\left({{\varepsilon}}{}_{{}}\right)+\frac{\tau^-}{\left|{\varepsilon}\right|}\mathcal{L}\left({{\varepsilon}}{}_{{}}\right) $$
	where $\ell\left( \cdot,\cdot \right)$ is the loss function and $ \tau^+$ and  $ \tau^-$ are the sampling frequency of  positive and negative edges, respectively. Given an edge $e$, Since the positive edge type element and negative edge type element of $l(e)$ can be expressed as ${{w}^{+/-}}^{T}f\left({e}\right)=\left \|  {w^{+/-}}  \right \| _2\:\left \|   f\left(e\right)   \right \| _2\cos(\theta) $, the partial derivative can be formulated as follows:
	
	\begin{align}
		&\frac{\partial\ell\left(e,{{y}}{}_{{}}\right)}{\partial{\left\|w^{+/-}\right\|}_2}=\frac{\partial\ell\left(e,{{y}}{}_{{}}\right)}{\partial l(e)}\frac{\partial l(e)}{\partial{\left\|w^{+/-}\right\|}_2}\\&= \begin{cases}{r^{{+/-}}}\left(e\right)\left\|f\left(e\right)\right\|_2\cos\left({{\theta}}_{{e}}^{+/-}\right),if\:y\ne {+/-} \\ {\left(r^{+/-}\right.}\left(e\right)-1)\left\|f\left(e\right)\right\|_2\cos\left({{\theta}}_{{e}}^{+/-}\right)\end{cases}
	\end{align}

	where  ${r^+}\left( e \right)$ and ${r^-}\left( e \right)$ represent the probability of being a positive edge and negative edge respectively, and ${{\theta}}_{{e}}^{+}$ is the angle between $f(e)$ and $w^+$ and  ${{\theta}}_{e}^{-}$ is the angle between $f(e)$ and $w^-$. Since  $\mathcal{L}\left(\varepsilon\right)$ is minimized during training, for $y=k$, $\cos\left({{\theta}}_{{e}}^{+}\right)$ is more likely to be positive, while $\mathcal{L}\left( \varepsilon\right)/\partial{\left\|w_+\right\|}_2 $ is more likely to be negative, thus raising ${\left \|w^{+} \right \|}_2$; on the other hand , $\cos\left({{\theta}}_{e}^{{-}}\right)$ is more likely to be negative, while $\mathcal{L}\left(\varepsilon\right)/\partial{\left\|w^-\right\|}_2 $ is more likely to be positive, resulting in a decrease in ${\left \|w^{-} \right \|}_2$. It suggests that the higher the sampling frequency, the larger the corresponding weight vector paradigm will be, which affects the decision boundary. 
	We rewrite the decision boundary as the following hyperplane:
	$$ \mathcal{B}\left(i,j\right)=\left\{{\mathit{e}\in\:\varepsilon}\middle|{\left\|{w}_i\right\|_2\:\cos}\left({{\theta}}_{{e}}^{{i}}\right)={\left\|{w}_j\right\|_2\:\cos}\left({{\theta}}_{{e}}^{{j}}\right)\right\} $$
	For a given graph, the direction of $w^+$ and $w^-$ is fixed after training a SGNN. As $\tau^+$ is usually larger than that of $\tau^-$, the paradigm $\small {\left \| w^{+} \right \|}_2 > {\left \| w^{-} \right \|}_2$ and the angle with the decision boundary $\small{{\theta}}^{{+}} >{{\theta}}^{{-}}$. It indicates that the decision boundary tends to be on negative edges, resulting in a smaller volume of feature space being allocated to negative edges. The edge regulator proposed in this paper tries to balance $\tau^+$ and  $\tau^-$ during graph augmentation, which can potentially adjust the decision boundary so as to alleviate the semantic unbalance problem in signed graphs.
	
	\section{B Dataset}
	
	\begin{longtblr}[
		label = none,
		entry = none,
		]{
			width = \linewidth,
			colspec = {Q[202]Q[155]Q[165]Q[183]Q[250]},
			cells = {c},
			hline{1-2,7} = {-}{},
		}
		Dateset        & \#Node & \#Pos Edges & \#Neg Edges & \#Ratio of Edges \\
		\footnotesize Bitcoin\_OTC   & 5,881   & 18,281       & 3,153        & 17.24\%          \\
		\footnotesize Bitcoin\_Alpha & 3,783   & 12,769       & 1,312        & 10.27\%          \\
		\footnotesize Congress       & 219    & 413         & 107         & 25.91\%          \\
		\footnotesize Wiki\_Election & 7,115   & 78,439       & 22,253       & 28.37\%          \\
		\footnotesize Chess          & 7,301   & 49,828       & 15,224       & 30.55\%          
	\end{longtblr}
	
	\begin{itemize}
		\item \textbf{Bitcoin\_OTC} \footnote{\url{https://www.bitcoin-otc.com/}} and \textbf{Bitcoin\_Alpha} \footnote{\url{http://www.btcalpha.com/}} are who-trusts-whom network of people who trade using Bitcoin on a platform called Bitcoin OTC. Since Bitcoin users are anonymous, there is a need to maintain a record of users' reputation to prevent transactions with fraudulent and risky users. 
		
		\item \textbf{Congress}\footnote{\url{https://github.com/NDS-VU/signed-network-datasets}} is a directed signed network collected as a part of a study published on 4/20/2020 for the International World Wide Web Conference. This specific network was sourced from congress votes, a directed signed network that represents politicians speaking in the United States Congress as nodes and mentions between speakers as directed edges. 
		
		\item \textbf{Wiki\_Election} \footnote{\url{https://www.wikipedia.org/}} is a directed signed network sourced from a study done by SNAP on the English Wikipedia website. This dataset represents a network of users that voted for or against each other in admin elections.
		
		\item \textbf{Chess}\footnote{\url{https://github.com/NDS-VU/signed-network-datasets}} is a directed signed network that is the result of chess games. Each node is a chess player, and a directed edge represents a game, with the white player having an outgoing edge and the black player having an ingoing edge.
	\end{itemize}
	\par In these datasets, users rate others from -10 (completely distrust)
	to 10 (completely trust). We treat the marks bigger than 0 as positive
	edges and others as negative edges.

	\section{C Experiment}

	\section{C.1 Backbones}
	\textbf{SGCN }is based on balance theory and generalizes GCN to signed graphs by aggregating and propagating messages through graph convolution to generate embeddings.
	
	\noindent \textbf{SNEA }is also based on balance theory and generalizes GAT to signed graphs by introducing an attention-based aggregator in the message passing process
	
	\noindent \textbf{RSGNN }quantifies the balance degree based on balance theory and proposes a new robust learning framework to enhance the stability of the SGNN model
	
	They have similar aggregation strategy, the aggregation for $L=1$ ($L$ is the aggregation layer) is defined as:
	
	$a_{i}^{pos(1)} = AGGREGATE  \left ( \left \{ h_{j}^{(0)}:v_{j}\in N_{i}^{+}    \right \}  \right )$
	
	$h_{i}^{pos(1)} = COMBINE  \left ( h_{i}^{(0)},a_{i}^{pos(1)} \right )$
	
	$a_{i}^{neg(1)} = AGGREGATE  \left ( \left \{ h_{j}^{(0)}:v_{j}\in N_{i}^{-}    \right \} \right )$
	
	$h_{i}^{neg(1)} = COMBINE  \left ( h_{i}^{(0)},a_{i}^{neg(1)} \right )$
	
	When $2\leq L \leq \omega$, the aggregation formula is as follows:
	
	\begin{equation}
		\begin{aligned} 
			a_{i}^{pos(L)} = AGGREGATE ( &\left \{ h_{j}^{pos(L-1)}:v_{j}\in N_{i}^{+} \right \} , \\& \left \{h_{j}^{neg(L-1)}:v_{j}\in N_{i}^{-} \right \}  )
		\end{aligned} 
	\end{equation}
	
	$h_{i}^{pos(L)} = COMBINE \left (\left \{ h_{i}^{pos(L-1)},a_{i}^{pos(L)}    \right \} \right )$
	
	\begin{equation}
		\begin{aligned} 
			a_{i}^{neg(L)} = AGGREGATE ( &\left \{ h_{j}^{neg(L-1)}:v_{j}\in N_{i}^{+} \right \} , \\& \left \{h_{j}^{neg(l-1)}:v_{j}\in N_{i}^{-} \right \} )
		\end{aligned} 
	\end{equation}
	
	$h_{i}^{neg(L)} = COMBINE\left (\left \{ h_{i}^{neg(L-1)},a_{i}^{neg(L)}    \right \} \right )$

	\clearpage
	
	\section{C.2 Experiment Result}

	\begin{figure*}
		\begin{longtblr}[
			label = none,
			entry = none,
			]{
				width = \linewidth,
				colspec = {Q[87]Q[146]Q[137]Q[148]Q[137]Q[137]Q[144]},
				row{1} = {c},
				cell{1}{1} = {c=2}{0.232\linewidth},
				cell{2}{1} = {r=4}{c},
				cell{2}{3} = {c},
				cell{2}{4} = {c},
				cell{2}{5} = {c},
				cell{2}{6} = {c},
				cell{2}{7} = {c},
				cell{3}{3} = {c},
				cell{3}{4} = {c},
				cell{3}{5} = {c},
				cell{3}{6} = {c},
				cell{3}{7} = {c},
				cell{4}{3} = {c},
				cell{4}{4} = {c},
				cell{4}{5} = {c},
				cell{4}{6} = {c},
				cell{4}{7} = {c},
				cell{5}{3} = {c},
				cell{5}{4} = {c},
				cell{5}{5} = {c},
				cell{5}{6} = {c},
				cell{5}{7} = {c},
				cell{6}{1} = {r=4}{c},
				cell{6}{3} = {c},
				cell{6}{4} = {c},
				cell{6}{5} = {c},
				cell{6}{6} = {c},
				cell{6}{7} = {c},
				cell{7}{3} = {c},
				cell{7}{4} = {c},
				cell{7}{5} = {c},
				cell{7}{6} = {c},
				cell{7}{7} = {c},
				cell{8}{3} = {c},
				cell{8}{4} = {c},
				cell{8}{5} = {c},
				cell{8}{6} = {c},
				cell{8}{7} = {c},
				cell{9}{3} = {c},
				cell{9}{4} = {c},
				cell{9}{5} = {c},
				cell{9}{6} = {c},
				cell{9}{7} = {c},
				cell{10}{1} = {r=4}{c},
				cell{10}{3} = {c},
				cell{10}{4} = {c},
				cell{10}{5} = {c},
				cell{10}{6} = {c},
				cell{10}{7} = {c},
				cell{11}{3} = {c},
				cell{11}{4} = {c},
				cell{11}{5} = {c},
				cell{11}{6} = {c},
				cell{11}{7} = {c},
				cell{12}{3} = {c},
				cell{12}{4} = {c},
				cell{12}{5} = {c},
				cell{12}{6} = {c},
				cell{12}{7} = {c},
				cell{13}{3} = {c},
				cell{13}{4} = {c},
				cell{13}{5} = {c},
				cell{13}{6} = {c},
				cell{13}{7} = {c},
				hline{1-2,6,10,14} = {-}{},
			}
			Method &            & Bitcoin\_OTC & Bitcoin\_Alpha & Congress    & Chess       & Wiki\_Election \\
			\small SGCN &\small Original     & \underline{0.818±0.009}  & 0.700±0.002     & 0.596±0.015 & 0.573±0.011 & \underline{0.778±0.012}    \\
			& \small DropMessage  & 0.809±0.008  & \underline{0.739±0.010}    & \underline{0.727±0.032} & 0.648±0.017 & 0.760±0.004    \\
			&\small GAug         & 0.637±0.047  & 0.641±0.004    & 0.723±0.062 & \underline{0.671±0.021} & 0.718±0.005    \\
			&\small SiGAug(Ours)         & \pmb{0.920±0.009}  & \pmb{0.945±0.002}    & \pmb{0.863±0.030} & \pmb{0.827±0.008} & \pmb{0.860±0.007}    \\
			\small SNEA &\small Original     & 0.930±0.004  & \underline{0.934±0.004}    & 0.858±0.011 & 0.471±0.004 & \underline{0.885±0.002}    \\
			& \small DropMessage  & \underline{0.934±0.001}  & 0.931±0.002    & \underline{0.868±0.007} & \underline{0.684±0.001} & 0.883±0.001    \\
			& \small GAug         & 0.648±0.007  & 0.694±0.005    & 0.842±0.008 & 0.165±0.052 & 0.791±0.001    \\
			& \small SiGAug(Ours)         & \pmb{0.940±0.009}  & \pmb{0.957±0.007}    & \pmb{0.884±0.010} &\pmb{0.819±0.019} & \pmb{0.905±0.003}    \\
			\small RSGNN &\small Original    & \underline{0.921±0.005}  & \underline{0.915±0.003}    & \underline{0.755±0.021} & 0.610±0.015        & \underline{0.866±0.005}    \\
			& \small  DropMessage & 0.603±0.001  & 0.613±0.008    & 0.600±0.005 & \underline{0.693±0.009} & 0.615±0.031    \\
			& \small GAug        & 0.644±0.021  & 0.722±0.017    & 0.570±0.001 & 0.379±0.002 & 0.793±0.001    \\
			&\small SiGAug(Ours)        & \pmb{0.939±0.004}  & \pmb{0.945±0.003}    & \pmb{0.845±0.061} & \pmb{0.867±0.003}            &  \pmb{0.890±0.002}          
		\end{longtblr}
		\centering\textbf{Table 3:}~~Link sign prediction performance on F1.
	\end{figure*}
	
	\begin{figure*}[]
		\begin{longtblr}[
			label = none,
			entry = none,
			]{
				width = \linewidth,
				colspec = {Q[162]Q[148]Q[160]Q[148]Q[148]Q[156]},
				column{1} = {c},
				cell{1}{2} = {c},
				cell{1}{3} = {c},
				cell{1}{4} = {c},
				cell{1}{5} = {c},
				cell{1}{6} = {c},
				cell{2}{2} = {c},
				cell{2}{3} = {c},
				cell{2}{4} = {c},
				cell{2}{5} = {c},
				cell{2}{6} = {c},
				cell{3}{2} = {c},
				cell{3}{3} = {c},
				cell{3}{4} = {c},
				cell{3}{5} = {c},
				cell{3}{6} = {c},
				cell{4}{2} = {c},
				cell{4}{3} = {c},
				cell{4}{4} = {c},
				cell{4}{5} = {c},
				cell{4}{6} = {c},
				cell{5}{2} = {c},
				cell{5}{3} = {c},
				cell{5}{4} = {c},
				cell{5}{5} = {c},
				cell{5}{6} = {c},
				hline{1-2,6} = {-}{},
			}
			Method       & Bitcoin\_OTC & Bitcoin\_Alpha & Congress    & Chess       & Wiki\_Election \\
			\small Original     & 0.332±0.012  & 0.154±0.008    & 0.252±0.043 & 0.301±0.008 & 0.397±0.016 \\
			\small DropMessage  & 0.331±0.004  & 0.162±0.017    & 0.242±0.037 & 0.304±0.009 & 0.395±0.022\\
			\small GAug         & 0.495±0.008  & 0.487±0.015    & 0.263±0.077 & 0.462±0.004 & 0.495±0.021\\
			\small   SiGAug(Ours)         &\pmb{0.991±0.002}  & \pmb{0.932±0.007}    & \pmb{0.766±0.024} & \pmb{0.981±0.003} & \pmb{0.960±0.008}
		\end{longtblr}
		\centering\textbf{Table 4}~Negative link prediction performance on precision.
	\end{figure*}
	
	\begin{figure*}[]
		\begin{longtblr}[
			label = none,
			entry = none,
			]{
				width = \linewidth,
				colspec = {Q[162]Q[148]Q[160]Q[148]Q[148]Q[156]},
				column{1} = {c},
				cell{1}{2} = {c},
				cell{1}{3} = {c},
				cell{1}{4} = {c},
				cell{1}{5} = {c},
				cell{1}{6} = {c},
				cell{2}{2} = {c},
				cell{2}{3} = {c},
				cell{2}{4} = {c},
				cell{2}{5} = {c},
				cell{2}{6} = {c},
				cell{3}{2} = {c},
				cell{3}{3} = {c},
				cell{3}{4} = {c},
				cell{3}{5} = {c},
				cell{3}{6} = {c},
				cell{4}{2} = {c},
				cell{4}{3} = {c},
				cell{4}{4} = {c},
				cell{4}{5} = {c},
				cell{4}{6} = {c},
				cell{5}{2} = {c},
				cell{5}{3} = {c},
				cell{5}{4} = {c},
				cell{5}{5} = {c},
				cell{5}{6} = {c},
				hline{1-2,6} = {-}{},
			}
			Method       & Bitcoin\_OTC & Bitcoin\_Alpha & Congress    & Chess       & Wiki\_Election \\
			\small Original     & 0.744±0.005  & 0.732±0.011    & 0.555±0.034 & \pmb{0.794±0.016} & 0.752±0.012 \\
			\small DropMessage  & 0.792±0.009  & 0.765±0.011    & 0.552±0.047 & 0.773±0.018 & 0.733±0.010\\
			\small GAug         & 0.651±0.013  & 0.690±0.017    & 0.55±0.039 & 0.568±0.012 & 0.622±0.018\\
			\small   SiGAug(Ours)         &\pmb{0.806±0.009}  & \pmb{0.882±0.012}    & \pmb{0.725±0.020} & 0.735±0.014 & \pmb{0.759±0.016}
		\end{longtblr}
		\centering\textbf{Table 5}~~Negative link prediction performance on recall.
	\end{figure*}
	
	\begin{figure*}[]
		\centering
		\includegraphics[width=1\linewidth]{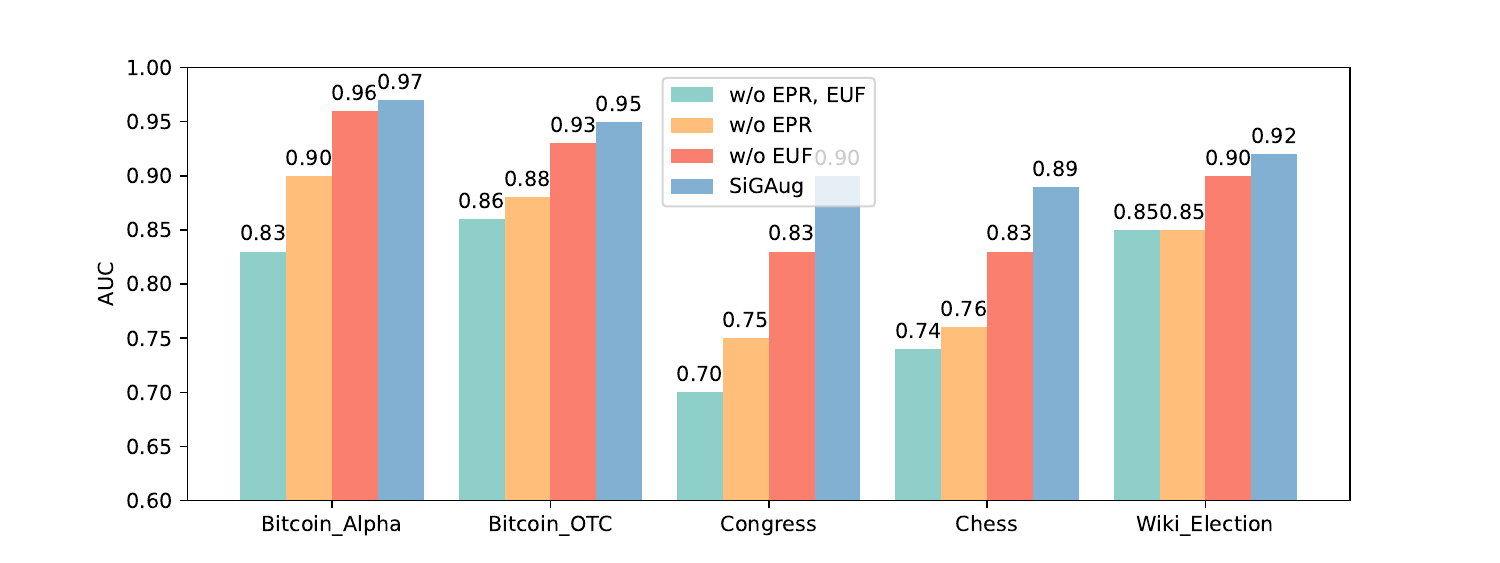}
		\centering{\textbf{Figure 7:}}~~AUC results of ablation study using different modules.
	\end{figure*}
	
	\begin{figure*}[t]
		\centering
		\includegraphics[width=1\linewidth]{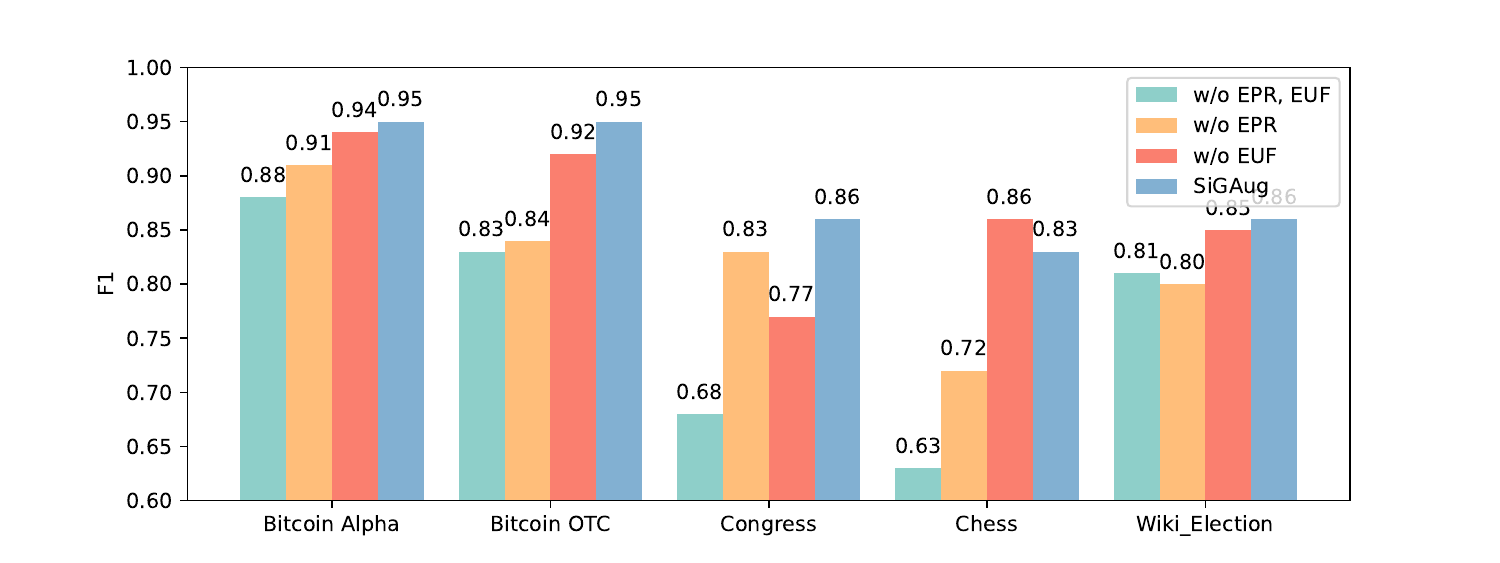}
		\centering{\textbf{Figure 8:}}~~F1 results of ablation study using different modules.
	\end{figure*}

	\begin{figure*}[H]
		\centering 
		\setcounter{subfigure}{0}
		\subfigure[$\mu$-AUC,  $\vartheta$=1/9, $\delta$=0.6]{
			\begin{minipage}[b]{0.40\linewidth}
				\begin{tikzpicture}[scale = 0.4] 
					\begin{axis}[
						xmin=0, 
						xmax=0.9,    
						ymin=0.4, 
						ymax=1,
						enlarge x limits=0.1,  
						xlabel=$\mu$, 
						ylabel=AUC, 
						xtick={0,0.1,0.3,0.5,0.7,0.9},     
						tick align=outside, 
						legend style={at={(1.1,1.3)},anchor=north,legend columns=-1} 
						]
						
						\addplot[,mark=*,violet] plot coordinates { 
							(0,0.7962)
							(0.1,0.8084)
							(0.3,0.8118)
							(0.5,0.8102)
							(0.7,0.8112)
							(0.9,0.8350)
						};
						
						\addlegendentry{\huge Bitcoin\_Alpha}
						
						
						\addplot[,mark=+,black] plot coordinates {
							(0,0.811)
							(0.1,0.819)
							(0.3,0.811)
							(0.5,0.813)
							(0.7,0.811)
							(0.9,0.860)
						};
						\addlegendentry{\huge Wiki\_election}
						
						\addplot[,mark=x,orange] plot coordinates {
							(0,0.590)
							(0.1,0.584)
							(0.3,0.590)
							(0.5,0.585)
							(0.7,0.639)
							(0.9,0.671)
						};
						\addlegendentry{\huge Chess}
						
						\addplot[,mark=triangle,cyan] plot coordinates {
							(0,0.595)
							(0.1,0.586)
							(0.3,0.560)
							(0.5,0.533)
							(0.7,0.575)
							(0.9,0.615)
						};
						\addlegendentry{ \huge Congress}
						
						\addplot[,mark=o,lime] plot coordinates {
							(0,0.807)
							(0.1,0.81)
							(0.3,0.816)
							(0.5,0.813)
							(0.7,0.812)
							(0.9,0.828)
						};
						\addlegendentry{\huge Bitcoin\_OTC}
					\end{axis}
				\end{tikzpicture}
			\end{minipage}
			\label{SNEA}
		}
		\subfigure[$\vartheta$-AUC,  $\mu$=0.7, $\delta$=0.6]{
			\begin{minipage}[b]{0.40\linewidth}
				\begin{tikzpicture}[scale =0.4] 
					\begin{axis}[
						xmin=0.2, 
						xmax=1,    
						ymin=0.4, 
						ymax=1,
						enlarge x limits=0.1,  
						xlabel=$\vartheta$, 
						ylabel=AUC, 
						tick align=outside, 
						legend style={at={(0.5,-0.2)},anchor=north} 
						]
						
						\addplot[,mark=*,violet] plot coordinates { 
							(0.2,0.645)
							(0.4,0.663)
							(0.6,0.696)
							(0.8,0.775)
							(1.0,0.816)
						};
						
						
						
						\addplot[,mark=+,black] plot coordinates {
							(0.2,0.58)
							(0.4,0.6)
							(0.6,0.69)
							(0.8,0.754)
							(1.0,0.83)
						};
						
						\addplot[,mark=x,orange] plot coordinates {
							(0.2,0.508)
							(0.4,0.537)
							(0.6,0.555)
							(0.8,0.575)
							(1.0,0.615)
						};
						
						\addplot[,mark=triangle,cyan] plot coordinates {
							(0.2,0.577)
							(0.4,0.525)
							(0.6,0.545)
							(0.8,0.600)
							(1.0,0.580)
						};
						
						\addplot[,mark=o,lime] plot coordinates {
							(0.2,0.667)
							(0.4,0.688)
							(0.6,0.711)
							(0.8,0.765)
							(1.0,0.828)
						};
					\end{axis}
				\end{tikzpicture}
			\end{minipage}
			\label{fig:SGCN}
		}
		
		\centering\textbf{Figure 9:}~~ The AUC of SiGAug (with random augmentation) when $\mu$ and $\vartheta$ vary.
	\end{figure*}

	\clearpage
	\section{D Method}
	\renewcommand{\thealgorithm}{1} 
	\begin{algorithm}[]
		\caption{Augmenter in SiGAug} 
		\begin{algorithmic}[1] 
			\renewcommand{\algorithmicrequire}{\textbf{Input:}} 
			\Require positive adjacency matrix $\small \mathcal{A}^+$, negative adjacency matrix $\small \mathcal{A}^-$
			\renewcommand{\algorithmicrequire}{\textbf{Data:}}
			\Require the utility threshold $ \mu $, the ratio of perturbed positive and negative edges $\vartheta $, the ratio of perturbed edges to original edges $\delta$
			\Ensure  augmented adjacency matrices $\tilde{\mathcal{A}}^{+}$ and $\tilde{\mathcal{A}}^{-}$
			\Repeat 
			\Function{ADD}{$\tilde{\mathcal{A}}^{+}$, $ \tilde{\mathcal{A}}^{-}$} 
			\While{$\delta<\varDelta$ and $\vartheta > \rho$} 
			\If{$e_{ij} $  is POSITIVE}
			\If{$filter\left( e_{ij} \right)$} 
			\State 
			$\tilde{\mathcal{A}}^{-}_{ij}\longleftarrow1$
			\State
			$perturbance\longleftarrow perturbance+1$
			\Else{ $e_{ij} $ is NEGATIVE}
			\State 
			$\tilde{\mathcal{A}}^{+}_{ij}\longleftarrow0$
			\State $perturbance\longleftarrow perturbance+1$
			\EndIf
			\EndIf
			\EndWhile
			\EndFunction
			~\\
			\Function{REMOVED}{$\tilde{\mathcal{A}}^{+}$, $\tilde{\mathcal{A}}^{-}$} 
			\While{$\delta<\varDelta$ and $\vartheta > \rho$} 
			\If{$e_{ij} $  is POSITIVE}
			\If{$filter\left( e_{ij} \right)$} 
			\State 
			$\tilde{\mathcal{A}}^{-}_{ij}\longleftarrow0$
			\State
			$perturbance\longleftarrow perturbance+1$
			\Else{ $e_{ij} $ is NEGATIVE}
			\State 
			$\tilde{\mathcal{A}}^{+}_{ij}\longleftarrow0$
			\State $perturbance\longleftarrow perturbance+1$
			\EndIf
			\EndIf
			\EndWhile
			\EndFunction
			
			\Until{$\delta=\varDelta $,  $\vartheta=\rho$} 
			\State \Return {$\tilde{\mathcal{A}}^{+}_{}$, ${\tilde{\mathcal{A}}^{-}_{}}$}
		\end{algorithmic}
	\end{algorithm}
	
	\renewcommand{\thealgorithm}{2} 
	\begin{algorithm}[htb]
		\caption{Fusing strategy in the augmenter} 
		\begin{algorithmic}[1] 
			\renewcommand{\algorithmicrequire}{\textbf{Input:}} 
			\Require positive adjacency matrix $ \tilde{\mathcal{A}^+}$, positive adjacency matrix $\tilde{\mathcal{A}^-}$
			\renewcommand{\algorithmicrequire}{\textbf{Data:}}
			\Require positive edge probability matrix $ \mathcal{M}^{+}$, negative edge probability matrix $ \mathcal{M}^{-} $
			\Ensure  augmented adjacency matrix $\tilde{\mathcal{A}}$
			\For{each $\tilde{\mathcal{A}}^{+}_{ij}$ and $\tilde{\mathcal{A}}^{-}_{ij}$}
			\If{$\tilde{\mathcal{A}}^{+}_{ij}$ and $\tilde{\mathcal{A}}^{-}_{ij}$ are 0}        
			\State $\tilde{\mathcal{A}}^{}_{ij}\longleftarrow 0$    
			\ElsIf{$\tilde{\mathcal{A}}^{+}_{ij}$ or $\tilde{\mathcal{A}}^{-}_{ij}$ is 0} 
			\State $\tilde{\mathcal{A}}^{}_{ij}\longleftarrow \tilde{\mathcal{A}}^{+}_{ij} - \tilde{\mathcal{A}}^{-}_{ij}$ 
			\Else{   
				\If{$ \mathcal{M}^{+} $ >$ \mathcal{M}^{-} $} 
				\State $\tilde{\mathcal{A}}^{}_{ij}\longleftarrow 1$  
				\Else{}
				\State $\tilde{\mathcal{A}}^{}_{ij}\longleftarrow -1$       
				\EndIf
			} 
			\EndIf
			\EndFor
			\State  \Return{$\tilde{\mathcal{A}}$} 
		\end{algorithmic}
	\end{algorithm}

\end{appendices}

\end{document}